\documentclass[fleqn,usenatbib]{mnras}

\usepackage{newtxtext,newtxmath}

\usepackage[T1]{fontenc}

\DeclareRobustCommand{\VAN}[3]{#2}
\let\VANthebibliography\thebibliography
\def\thebibliography{\DeclareRobustCommand{\VAN}[3]{##3}\VANthebibliography}

\usepackage{graphicx}	
\usepackage{amsmath}	
\usepackage{subcaption}

\newcommand{\eazy}{{\tt{EAZY}}}

\newcommand{\sextractor}{{\tt{SExtractor}}}
\newcommand{\galfit}{{\tt{GALFIT}}}
\newcommand{\imfit}{{\tt{IMFIT}}}
\newcommand{\webbpsf}{{\tt{WebbPSF}}}

\newcommand{\logm}{$\log \left(M_* / M_{\odot}\right)$}

\title[EPOCHS VI]{EPOCHS VI: The Size and Shape Evolution of Galaxies since $z \sim 8$ with JWST Observations}

\author[K. Ormerod et al.]{
K. Ormerod,$^{1}$\thanks{E-mail: katherineeormerod@gmail.com}
C. J. Conselice,$^{1}$
N. J. Adams,$^{1}$
T. Harvey,$^{1}$
D. Austin,$^{1}$
J. Trussler,$^{1}$
L. Ferreira$^{2}$, \newauthor
J. Caruana$^{3,4}$,
G. Lucatelli$^{1}$,
Q. Li$^{1}$,
W. J. Roper$^{5}$\\ 
\\
$^{1}$Jodrell Bank Centre for Astrophysics, University of Manchester, Oxford Road, Manchester M13 9PL, UK\\
$^{2}$Department of Physics \& Astronomy, University of Victoria, Finnerty Road, Victoria, British Columbia, V8P 1A1, Canada \\
$^{3}$ Department of Physics, University of Malta, Msida MSD 2080, Malta \\
$^{4}$ Institute of Space Sciences \& Astronomy, University of Malta, Msida MSD 2080, Malta \\
$^{5}$Astronomy Centre, University of Sussex, Falmer, Brighton BN1 9QH, UK\\
}

\date{Accepted XXX. Received YYY; in original form ZZZ}

\pubyear{2023}

\begin{document}
\label{firstpage}
\pagerange{\pageref{firstpage}--\pageref{lastpage}}
\maketitle

\begin{abstract}
We present the results of a size and structural analysis of 1395 galaxies at $0.5 \leq z \lesssim 8$ with stellar masses \logm $>$ 9.5 within the \emph{JWST} Public CEERS field that overlaps with the HST CANDELS EGS observations. We use \galfit\ to fit single Sérsic models to the rest-frame optical profile of our galaxies, which is a mass-selected sample complete to our redshift and mass limit.  Our primary result is that at fixed rest-frame wavelength and stellar mass, galaxies get progressively smaller, evolving as $\sim (1+z)^{-0.71\pm0.19}$ up to $z \sim 8$.   We discover that the vast majority of massive galaxies at high redshifts have low Sérsic indices, thus do not contain steep, concentrated light profiles.  Additionally, we explore the evolution of the size-stellar mass relationship, finding a correlation such that more massive systems are larger up to $z \sim 3$. This relationship breaks down at $z > 3$, where we find that galaxies are of similar sizes, regardless of their star formation rates and Sérsic index, varying little with mass.   We show that galaxies are more compact at redder wavelengths, independent of sSFR or stellar mass up to $z \sim 3$.  We demonstrate the size evolution of galaxies continues up to $z \sim 8$, showing that the process or causes for this evolution is active at early times.  We discuss these results in terms of ideas behind galaxy formation and evolution at early epochs, such as their importance in tracing processes driving size evolution, including minor mergers and AGN activity.

\end{abstract}

\begin{keywords}
galaxies: structure -- galaxies: evolution -- galaxies: high-redshift
\end{keywords}



\section{Introduction}

Ever since the discovery of galaxies, their extended nature has been a clear indicator of their different properties from unresolved sources such as stars. These resolved properties have been of interest for many years, and in many ways are the oldest studied properties of galaxies \citep[e.g.,][]{hubble1926,Buitrago2008, Conselice2014a, Ferreira22b, vdw2014, Suess2022, ferreira22a, Kartaltepe2022}. As far back as the 18th century, the Herschels catalogued what we now know to be extragalactic objects, commenting on their appearance \citep{herschel_1786}.  Following up on this, Lord Rosse discovered spiral structures in galaxies, through his observations of M51 and other nearby galaxies \citep{rosse_1850}. After inferring the distances to these objects, photography allowed Hubble and his immediate successors to develop the dominant morphological classification scheme we use today -- the Hubble Sequence -- which classifies extragalactic objects as spiral, elliptical, or irregular \citep{hubble1926}.  This has continued until this day, with James Webb Space Telescope (\emph{JWST}) observations setting us on a new pathway towards understanding the structures and morphologies of the very first galaxies \citep[e.g.,][]{Whitney2021,ferreira22a, Ferreira22b, Suess2022, Kartaltepe2022, huertas_company_2023, ono_2023, jacobs2023, tacchella_2023, morishita2023}.

The study of galaxy structure and morphology are amongst the oldest subjects within the extragalactic field, and continue to hold significant importance in our quest to understand the evolutionary processes of galaxies over cosmic time, however our knowledge about these features remains limited at $z > 3$. Tracking the changes in the structural properties of galaxies from the era of early galaxy formation until the present day provides valuable insights into the processes of galaxy evolution. There have been major efforts over the past 30 years to study morphology and structure of distant galaxies with the Hubble Space Telescope \citep[\emph{HST}; ][]{Buitrago2008, Conselice2014a, Delgado-Serrano2010}, where the rest-frame optical properties of galaxies up to $z \sim 3$ can be studied and examined. These HST observations have shown us that galaxies appear to become progressively more irregular and peculiar at higher redshifts \citep[e.g.,][]{Conselice2003a,Lotz2004, Mortlock2013, Delgado-Serrano2010, Schawinski2014, Conselice2014a, Whitney2021}.  

However, the red Hubble filters are limited, and do not allow us to measure or observe the rest-frame optical light of galaxies back to within the first few Gyr of the universe. In fact, the reddest HST filter, F160W on Hubble`s Wide Field Camera 3 (WFC3), only probes the rest-frame visible light of galaxies up to $z \sim 2.8$, but galaxies exist at much higher redshifts, and this paper aims to probe their evolution\citep[e.g.,][]{adams2023, austin_2023, curtis-lake_2023, Castellano_2022, Naidu_2022, Finkelstein_2023, Atek_2022, Yan_2022, Donnan_2022,Harikane_2023}. JWST is the best approach for measuring galaxies in the rest-frame optical where the effects of dust are limited.  We can see why this longer wavelength data is needed through the use of cosmological simulations. These simulations predict that at higher redshift, galaxies in the absence of the effects of dust are intrinsically more compact resulting in a negative far-UV size-luminosity relation. However, when simulated with dust, the bright cores of these galaxies are attenuated, increasing the observed half-light radius. This brings the FUV size-luminosity relations in line with observations. Simulations, in agreement with observations, have also shown that galaxy sizes typically increase with increasing stellar mass, and show that compact galaxies may grow in size due to mergers or renewed star formation, with high-redshift stars moving outwards \citep[e.g.,][]{furlong_2017, marshall_2022, Roper_2022}.

The relatively recently launched James Webb Space Telescope allows us to obtain the same type of data with the Near Infrared Camera (NIRCam) probing rest frame optical light as far out as $z\sim9$, with filters reaching up to $\sim4.4\mu$m. The superior resolution of JWST and the extended wavelengths of its filters allow us to examine galaxy structure in significantly greater detail than with \emph{HST} \citep[][]{Ferreira22b}.   Early \emph{JWST} observations reveal morphological and structural features of galaxies at $1.5 < z < 8$ that were not possible to discern fully with \emph{HST}, thus resulting in the re-classification of many galaxies previously believed to have peculiar morphologies \citep[e.g.,][]{Ferreira2020}. A major discovery has been that galaxies appear morphologically much more disc-like than previously thought \citep[e.g.,.][]{ferreira22a}. While these early \emph{JWST} papers show that galaxies at $z > 2$ are different than we thought on the basis of \emph{HST} imaging, a significant amount of quantitative analysis is still needed.

As such, in this paper we present an analysis of the sizes and Sérsic indices of a mass complete sample of massive galaxies with stellar masses \logm $> 9.5$, for which we can now quantitatively measure rest-frame structure up to $z \sim 8$ with \emph{JWST}. Whilst previously we could also measure galaxy sizes and morphologies with \emph{HST}, these are often unreliable due to image fidelity and the range of wavelengths we were able to probe \citep[][]{Ferreira22b}. Our results at $0.5 < z < 8$ allow us to probe deeper and at higher redshifts than previous studies. 

We present a quantitative analysis of measured galaxy shapes, based on the Sérsic index, $n$, obtained from Sérsic profile fitting, and size measurements, based on half-light radii measurements, to determine the evolution of galaxy structure over most of cosmic time. Whilst the Sérsic profile is one of the simplest forms that can be fit to a light profile, this inital analysis is important as it enables a continuation of work completed with HST, and allows a direct comparison of the results at different redshift ranges. It should be noted that a Sérsic profile will not model the full detail of some galaxies, although the galaxies within our sample are not resolved enough to see the very core regions or to show kinks that are seen in surface brightness profiles of very nearby galaxies. In this paper we present an intial analysis of Sérsic profile fitting with JWST, to answer how galaxy sizes and their overall shapes change for a mass-complete sample,and complete further analysis on sub-samples based upon two properties: specific star formation rate, and Sérsic indices. We separate our sample into sub-samples based on sSFR to investigate the evolution of Sérsic index in the more star-forming and passive populations. We also separate our sample into sub-samples with a high and low Sérsic index, with a separating value of $n = 2$, as a proxy for elliptical-like and disc-like galaxies, and to investigate their sizes across cosmic time.  This differs from previous work which has focused on either just the highest redshift galaxies \citep[e.g.,][]{ono_2023} or those which are passive \citep[e.g.,][]{ito_2023}. The evolution of these properties within a stellar mass selection is a key observable for galaxy evolution as well as an important way to trace processes that drive galaxy size evolution in massive galaxies, including galaxy minor mergers and AGN activity \cite[e.g.,][]{Bluck2012}.

The structure of this paper is as follows: Section~\ref{sec:data} discusses the data used and the data reduction process, Section~\ref{sec:properties} discusses properties of the galaxies within our sample. The fitting process is explained in Section~\ref{sec:galfit}, and we discuss how we select a sample of robust morphological fits in Section~\ref{sec:sample_selection}. We present our results, along with a comparison to simulations to confirm that our findings are not the result of redshift effects in Section~\ref{sec:results}. 
We assume a standard $\Lambda$~CDM cosmology throughout of $\Omega_m = 0.3$, $\Omega_\Lambda = 0.7$, and $H_0 = 70$ km s$^{-1}$Mpc$^{-1}$. Where we reference galaxy `sizes', we are referring to the half-light radii of our objects. All magnitudes are given in the AB system \citep{oke, oke_gunn}.

\section{Data}
\label{sec:data}
We use  \emph{JWST} NIRCam imaging \citep{nircam} to analyse the light profiles of a large sample of high-redshift galaxies in the F115W, F150W, and F200W short-wavelength (SW) bands, and F277W, F356W, F410M, and F444W long-wavelength (LW) bands.
The Cosmic Evolution Early Release Science (CEERS, PID: 1345, PI: S. Finkelstein) Survey \citep{ceersprop, ceers1, Bagley_2023} is one of 13 \emph{JWST} ERS programmes, with the goal of examining galaxy formation at 0.5 < z < 10 and perhaps beyond. The galaxies analysed in this work are within the CEERS NIRCam footprint, and within the Cosmic Assembly Near-infrared Deep Extragalactic Legacy Survey (CANDELS) observations \citep{koekemoer2011}. This is important as it allows us to use both the \emph{JWST} data as well as the deep data from \emph{HST's} WFC3 and Advanced Camera for Surveys (ACS), which also aids the determination of photometric redshifts. As such, the photometric redshifts, stellar masses, and star formation rates used in this paper are based on the original CANDELS+GOODS WFC3/ACS imaging and data, Spitzer/IRAC S-CANDELS observations \citep{ashby2015}, and Canada-France-Hawaii Telescope (CFHT) ground-based observations \citep{stefanon2017}. A summary of \emph{HST} filters used and their $5\sigma$ depths is shown in \autoref{tab:depths}.

\begin{table}
    \centering
    \begin{tabular}{c|c} 
     Filter Name    & Depth  \\ \hline
         \emph{HST}/ACS F606W& 28.8 \\
         \emph{HST}/ACS F814W & 28.2 \\
         \emph{HST}/WFC3 F125W & 27.6 \\
          \emph{HST}/WFC3 F140W&26.8 \\
           \emph{HST}/WFC3 F160W&27.6 \\
    \end{tabular}
    \caption{The $5\sigma$ depths of the \emph{HST} photometric data covering the EGS, for full details see \citet{stefanon2017}.}
    \label{tab:depths}
\end{table}

The CANDELS survey was designed to investigate galaxy evolution and the birth of black holes at $1.5 < z < 8$, and consists of multi-wavelength observations in five fields. 
The CANDELS/DEEP survey covers 125 arcmin$^{2}$ within GOODS-N and GOODS-S, and the remainder consists of the CANDELS/Wide survey, covering three additional fields (Extended Groth Strip, COSMOS, Ultra-Deep Survey), covering a total of 800 arcmin$^{2}$ across all fields \citep{CANDELS}.   The primary sample we select from originates from these deep observations within the EGS, where there is overlap with the CEERS \emph{JWST} NIRCam data, and whose analysis is described in detail in \citet[][]{duncan2019}.

\subsection{CEERS Data Reduction}\label{subsec:reduction}

We process the \emph{JWST} data products on this field using a modified version of the official \emph{JWST} pipeline, explained in depth in \citet{adams2023, adams2023epochs}. We use the standard \emph{JWST} pipeline (pipeline version 1.8.2 and Calibration Reference Data System v0995), with some minor modifications. Between Stage 1 and Stage 2, we subtract templates of `wisp' artefacts from the F150W and F200W data \footnote{\url{https://jwst-docs.stsci.edu/jwst-near-infrared-camera/nircam-instrument-features-and-caveats/nircam-claws-and-wisps}}. After Stage 2 of the pipeline we apply a correction for 1/F noise, derived by Chris Willot.\footnote{\url{https://github.com/chriswillott/jwst}} We extract the sky subtraction step from Stage 3 of the pipeline and run this on each NIRCam frame independently. We then align calibrated imaging for each exposure to GAIA Data Release 3 \citep{Gaia}, using \texttt{tweakreg} from the \textsc{drizzlepac python}\footnote{\url{https://github.com/spacetelescope/drizzlepac}} package. We finally pixel-match the final mosaics using \texttt{astropy reproject}.\footnote{\url{https://reproject.readthedocs.io/en/stable/}} The final resolution of our drizzled images is 0.03 arcsec per pixel. The total unmasked area of \emph{JWST} images used in this paper is 64.15 square arcminutes, and the average depths of each filter are listed in \autoref{tab:CEERS_depths}.

\begin{table}
    \centering
    \begin{tabular}{c|c}
         Filter & Depth  \\ \hline
         F115W & 28.75 \\
         F150W & 28.60 \\
         F200W & 28.80 \\
         F277W & 28.95 \\
         F356W & 29.05 \\
         F410M & 28.35\\
         F444W & 28.60
    \end{tabular}
    \caption{Average $5\sigma$ depths in our reduced CEERS images, for point sources in 0.32'' diameter apertures. This aperture size is chosen as it encloses the central $70-80\%$ of the flux of a point source, and is small enough to avoid contamination. Depths are calculated by placing random apertures in regions of the image that are empty based on the final segmentation maps.}
    \label{tab:CEERS_depths}
\end{table}

\section{Galaxy Properties}
\label{sec:properties}

\subsection{Photometric Redshifts, Stellar Masses and Star Formation Rates} \label{subsec:redshift}
\begin{figure}
     \centering
     \includegraphics[width=0.95\columnwidth]{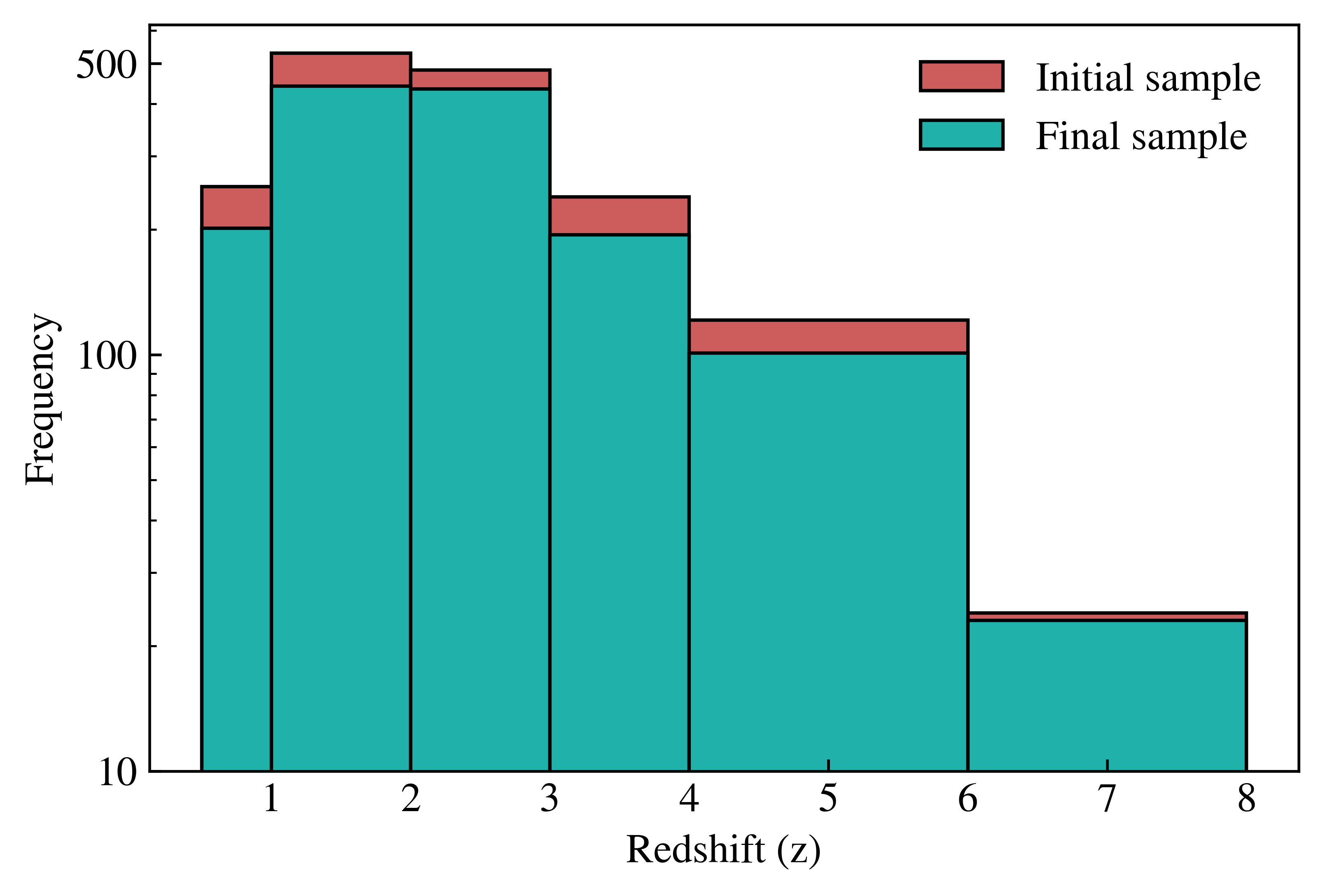}
     \caption{Histogram showing the distribution of photometric redshifts within our redshift bins. The `Initial sample' is the sample of 1649 high mass galaxies we begin the fitting process with, and the `Final sample' represents the galaxies that are selected as good fits in \autoref{sec:sample_selection}.}
     \label{fig:phot_z_hist}
 \end{figure}

 We begin our analysis with a catalogue of 1649 massive objects with \logm $ > 9.5$, which have photometric redshifts and physical properties calculated in previous works. The photometric redshifts used in this paper are calculated in \citet{duncan2019}, using the \eazy\ \citep{brammer_2008} photometric redshift code, with three separate template sets fitted to the observed photometry. These templates include zero-point offsets, which alter the input fluxes, and fix additional wavelength-dependent errors (see \citet{duncan_2018a, duncan_2018b} for full details). A Gaussian process code \citep[GPz;][]{almosallam_2016} is used to measure further empirical estimates, using a subset of the photometric bands. Individual redshift posteriors are calculated, and all measurements are combined in a statistical framework via a Bayesian combination to give a final estimate of redshift. From a comparison with spectroscopic redshifts, for sources where spectroscopic measurements are available, these photometric redshifts are seen to have a high degree of accuracy; for full details, see Section 2.4 of \citet{duncan2019}. We show the photometric redshift distribution of our sample in \autoref{fig:phot_z_hist} for both our initial and final samples. We note that there is a significant decrease in the number of galaxies in the highest redshift bin, although this is not unexpected due to the catalogue being compiled from previous HST works, thus HST dark galaxies are not included in this sample.

The stellar masses we use are those measured in \citet{duncan2014, duncan2019}, using a custom template fitting code (see Section 4 of \citet{duncan2014}). With this custom spectral energy distribution (SED) fitting code, the stellar mass is measured at all redshifts within the photometric redshift fitting range. The masses also have a `template error function', described in \citet{brammer_2008}, accounting for uncertainties driven by the template set and wavelength effects. These stellar mass measurements assume a BC03 stellar population synthesis (SPS) model \citep{bruzual_2003}, with a wide range of stellar population parameters, and a \citet{chabrier_2003} initial mass function (IMF). The star formation histories used within these fits follow the form $\mathrm{SFR} \propto \mathrm{e}^{-t / \tau}$, with timescales of $|\tau|$=$0.25, 0.5, 1, 2.5, 5, 10$, where negative values of $\tau$ represent exponentially increasing histories. A short burst model is also used ($\tau = 0.05$), as well as continuous star formation models ($\tau$ = $1$/$H_0$). Nebular emission is also included in the model SEDs assuming an escape fraction of $f_{esc} = 0.2$. In order to ensure that our stellar masses do not suffer from systematic biases, they are compared to stellar masses calculated independently within the CANDELS collaboration \citep{santini_2015}. While there is some scatter between the two mass estimates, there is no significant bias (see Section 2.5 of \citet{duncan2019} for a detailed discussion).  We aim to calculate masses using \emph{JWST} data for galaxies at $z > 4.5$ in Harvey et al., in prep.

Star formation rates for our sample of galaxies are calculated using the UV slope ($\beta$) of the spectral energy distribution, which gives a measure of the dust attenuation within the galaxy. We aim to measure this with \emph{JWST} within this EPOCHS paper series (Austin et al., in prep). From this, we correct for dust and obtain the total star formation rates for our galaxies, which agree well with star formation rates derived directly from SED fitting \citep{duncan2014, duncan2019}. We avoid degeneracies between the stellar masses and star formation rates by deriving the star formation rates using this separate method. The SFR and stellar mass estimates used in this work are calculated in \citet{duncan2014, duncan2019}, and are compared to several other works within the CANDELS collaboration which find no significant biases. The observed scatter found is due to the photo-z estimates used not being exactly the same in each paper, and is most significant at \logm $< 9$, which is not within the mass range of this paper.

\subsection{Visual Classifications}
\label{subsec:visual}
For 470 objects within our sample, we use visual classifications from \citet{ferreira22a} as part of our analysis. The categories we use are as follows:
\begin{enumerate}
    \item \textit{Discs:} Sources with a resolved disc with an outer area of lower surface brightness, that regularly increases towards the centre of the galaxy. 
    \item \textit{Spheroids:} Resolved sources that are symmetrical, with a centrally concentrated, smooth light profile, that are round or elliptical.
    \item \textit{Peculiar:} Resolved sources with a disturbed morphology, which dominates any smooth components.
    \item \textit{Other:} Mainly made up of sources classified as `ambiguous', due to the classifiers not reaching a majority agreement on the classification of a source. This category also contains sources that are classified as point sources due to an angular size smaller than the full width half maximum (FWHM) of the point spread function (PSF), or clear spikes that are consistent with point sources, and any sources that were unable to be classified due faintness or image issues.
\end{enumerate}

\section{Morphological Fitting}
\label{sec:galfit}

We use \galfit\ version 3.0.5 \citep{Galfit1, Galfit2} to fit a single Sérsic light profile to each galaxy.  Ultimately, the overall goal with \emph{JWST} is to measure the light profiles in more detail, such as obtaining bulge to disk ratios and other features \citep[][]{margalef-bentabol2016}. However, it is important to first determine structural properties using single profile fitting, and assess how they characterise the data \citep[e.g.,][]{vanderwal2012, vdw2014, Suess2022}.

\galfit\ is a least-squares-fitting algorithm which finds the optimum solution to the surface brightness profiles for galaxies through using a Levenberg-Marquardt algorithm. \galfit\ uses the reduced chi-squared, $\chi_{\nu}^{2}$, to determine goodness-of-fit and finds the best fit model through $\chi_{\nu}^2$ minimisation. The  $\chi_{\nu}^2$ is given by by:
\begin{equation}
\label{eqn:gal_chi}
\chi_\nu^2=\frac{1}{N_{\mathrm{DOF}}} \sum_{x=1}^{n x} \sum_{y=1}^{n y} \frac{\left(f_{\mathrm{data}}(x, y)-f_{\mathrm{model}}(x, y)\right)^2}{\sigma(x, y)^2}
\end{equation}
summed over $nx$ and $ny$ pixels, and where $N_{DOF}$ is the number of degrees of freedom. As seen in \autoref{eqn:gal_chi}, \galfit\ requires a data image from which the galaxy surface brightness is measured, $f_{\mathrm{data}}(x,y)$ and a sigma image, $\sigma(x,y)$, giving the relative error at each position within the image, which are then used to calculate the model image, $f_{\mathrm{model}}(x,y)$.

\begin{figure}
	\includegraphics[width=\columnwidth]{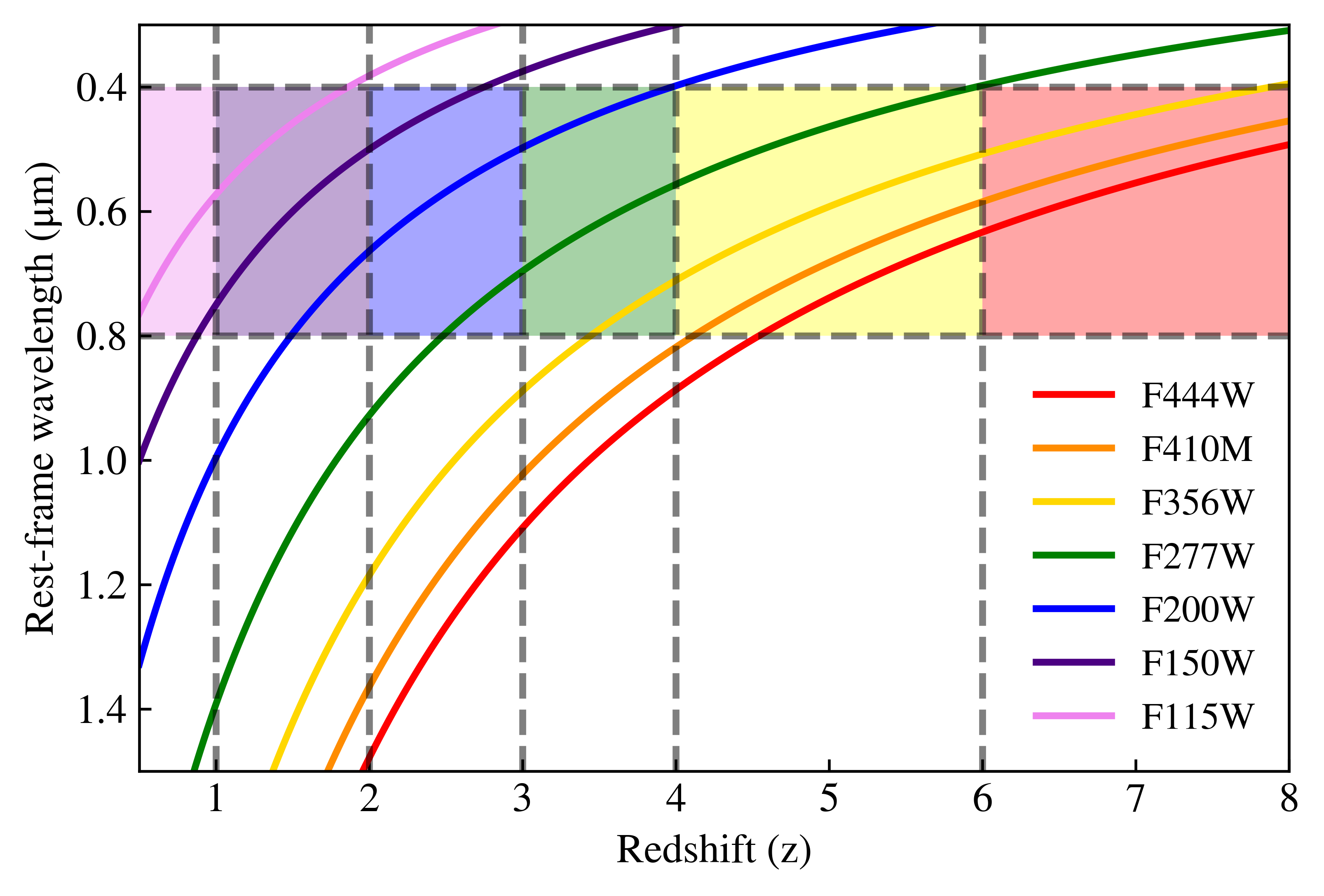}
    \caption{Plot showing the rest-frame wavelengths at given redshifts, for all filters used within the JWST CEERS NIRCam observations, and used within this paper. The shaded regions show the selected filter we use to observe sources in the rest-frame optical at the given redshift.  The grey dashed lines show where the filter we use to provide a rest-frame optical view of galaxies at different redshift changes.}
    \label{fig:filters}
\end{figure}

We run \galfit\ for all available filters, but only report results here for the filters that best match the rest-frame optical wavelength of the source. This minimises, or even eliminates, the effect of morphological \textit{k}-correction, as the qualitative and quantitative structure of galaxies changes as a function of wavelength \citep{TaylorMager2007}, which can result in significant structural changes between rest-frame UV and rest-frame optical images. This also enables us to limit the CMOD effect \citep{papaderos2023}, by probing rest-frame optical wavelengths, which is possible up to $z\sim8$ with JWST. The band selected at a given redshift is shown in \autoref{fig:filters}. The figure shows which filter we use within different redshift ranges, and what rest-frame wavelength we probe within that filter at that redshift. As can be seen, we are always probing the rest-frame optical at wavelengths redder than the Balmer break at all epochs in which we view our galaxy sample.

The Sérsic profile we use has the form 
\begin{equation}
    I(R)=I_e \exp \left\{-b_n\left[\left(\frac{R}{R_e}\right)^{1 / n}-1\right]\right\},
    \label{eqn:Sérsic}
\end{equation}
where $I(R)$ is the intensity at a distance $R$ from the centre of the galaxy, $R_{e}$ is the half-light radius of galaxy (the radius where 50\% of the total luminosity is enclosed), $I_e$ is the intensity at the half-light radius, $n$ is the Sérsic index, which controls the shape of the light profile of the galaxy \citep{Sersic, ciotti_1991, caon_1993}, and $b_{n}$ can be approximated as $b(n)\approx 2 n-\frac{1}{3}+\frac{4}{405 n}+\frac{46}{25515 n^2}$ \citep{ciotti1999}.  \galfit\ gives us a best fitting value for each of these terms.  The errors on these values are also calculated through this method, and a full description of the \galfit\ error calculation can be found in \citet{Galfit1}.

\subsection{\galfit\ Pipeline}
\label{subsec:pipeline}
\begin{figure}
	\includegraphics[width=\columnwidth]{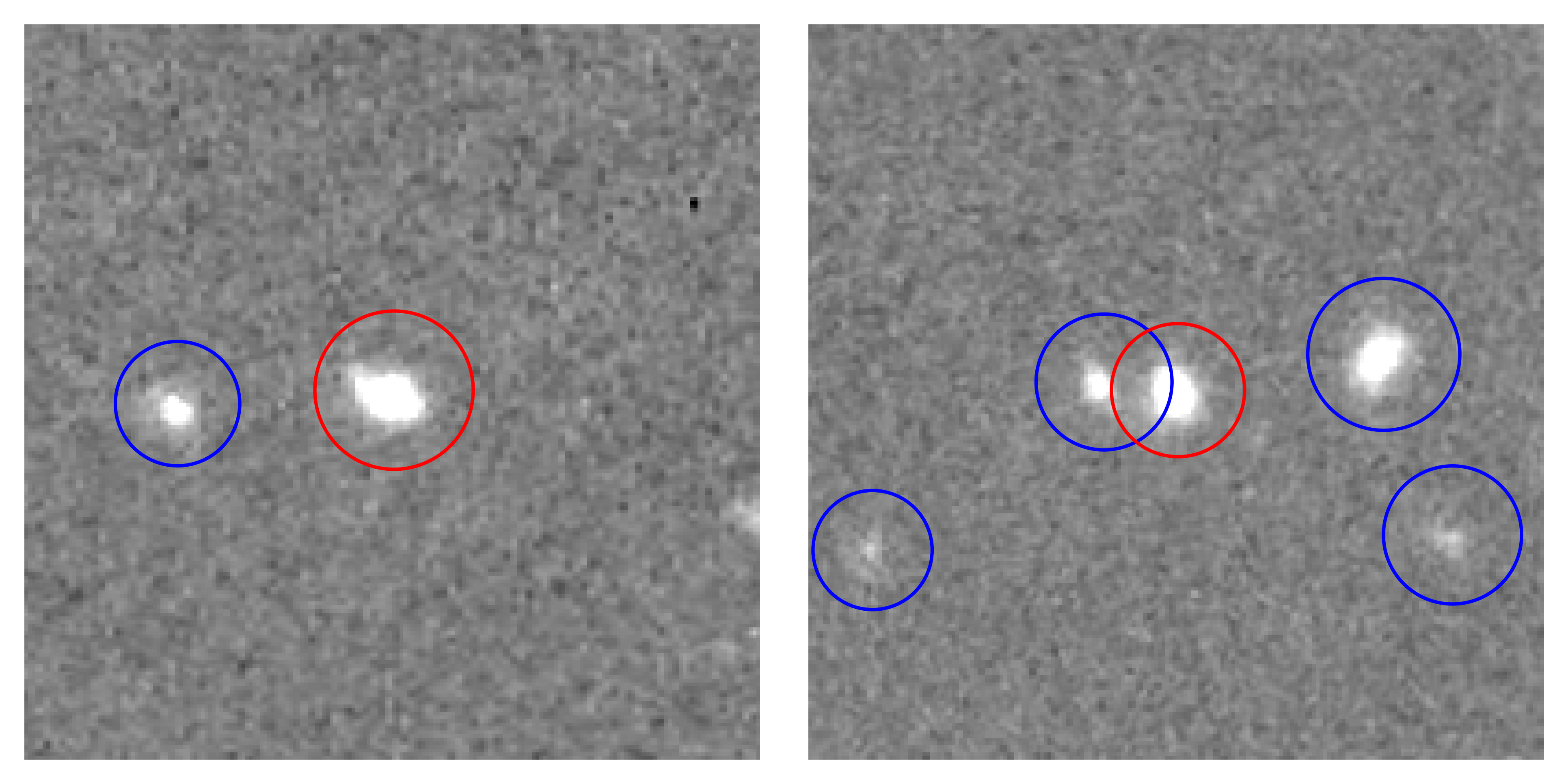}
    \caption{Plots showing the Kron radii (semi-major axis of the Kron ellipse), where the red radius in each image is that of the primary source, and the blue radii are those of the neighbouring sources.  We give several scenarios for how these systems would be found.  Left: No other sources would be fit simultaneously to the primary, and the pixels belonging to all other objects according to the \sextractor\ segmentation maps are masked. Right: The source where the Kron radius overlaps that of the primary source are  simultaneously fit in this instance, and all other sources would be masked. Cutout sizes shown are 6'' $\times$ 6''.}
    \label{fig:masks}
\end{figure}

We use a custom pipeline for single component Sérsic fits with \galfit. The process is as follows:
\begin{enumerate}
    \item \textit{Source Detection}: We use \sextractor\ \citep{bertin1996} to detect sources within the F444W images for each CEERS pointing, following the parameters and method in \citet{adams2023}. These catalogues are then cross-matched within 1 arcsecond to the catalogues created from the analysis completed in \citet{duncan2019} to create our final catalogues for each pointing. The average separation between both catalogues is $\sim 0.15''$

    \item \textit{Cutout and Mask Creation}: We create 200 x 200 pixel (6$'' \times 6''$) cutouts of each source, in order to ensure the entire surface brightness profile of the galaxy is enclosed within the cutout, along with that of any neighbours that may need to be modelled simultaneously. We use the \sextractor\ segmentation maps to make masks, creating the same 200 x 200 pixel cutout of the segmentation map, and then masking the necessary objects. 
    
In order to create masks for each object, and select which neighbouring objects must be masked, we use the Kron ellipses \citep{kron1980}, as defined by \sextractor\, and plot circular apertures with a radius equal to the semi-major axis of the Kron ellipse. We do this to select galaxies that are sufficiently close enough for their surface brightness profile to interfere with that of the target, `primary', object and must be fit simultaneously. If any neighbouring galaxy has an overlapping Kron aperture with that of the primary object, the neighbouring galaxy is deemed to be sufficiently close that it must be modelled alongside the primary galaxy, to account for both light profiles. We do this for as many neighbouring objects as necessary. Objects that do not have an overlapping Kron aperture are far enough away that they do not also need to be fit, and therefore are masked instead, primarily to save computational time. Through visual inspection of fits and residuals of fits from the data, we conclude that this criterion is good for determining when to fit neighbouring objects. The pixels that are masked are those that the \sextractor\ segmentation maps assigns to each source. An example of these selection criteria is shown in \autoref{fig:masks}.

    \galfit\ also requires a sigma image to give relative weight to the pixels in the image during the fit. As the input sigma image, we use the `ERR' extension of the images, which is a measure of the noise of the image, and this is created using the same method as the object cutout images.

    \item \textit{Input Parameters}: In order to fit a single Sérsic profile, initial estimates of parameters must be provided to \galfit. The input parameters that \galfit\ requires estimates for are $x$ and $y$ image coordinates, total magnitude, half-light radius, axis ratio, position angle, and Sérsic index. Similarly to \citet{Kartaltepe2022}, we use the \sextractor\ catalogue for our initial parameter estimates. As these catalogues do not contain values of Sérsic index, we estimate this as $n =1$ initially and allow the fitting process to determine the best fit Sérsic index. We find that using different values of $n$ as the initial estimate have virtually no effect on the output parameters. We only apply constraints to the image position of the sources within $\pm $ 2 pixels to ensure the correct source is being fit.

    \item \textit{Point Spread Function}: \galfit\ requires the appropriate PSF for each filter, which is obtained using \webbpsf\ \citep{perrin2014}, and resampled to our pixel scale. We experimented with different PSFs created through this method and find that the results do not significantly change.
\end{enumerate}

\begin{figure}
    \centering
    \includegraphics[width = 0.99\columnwidth]{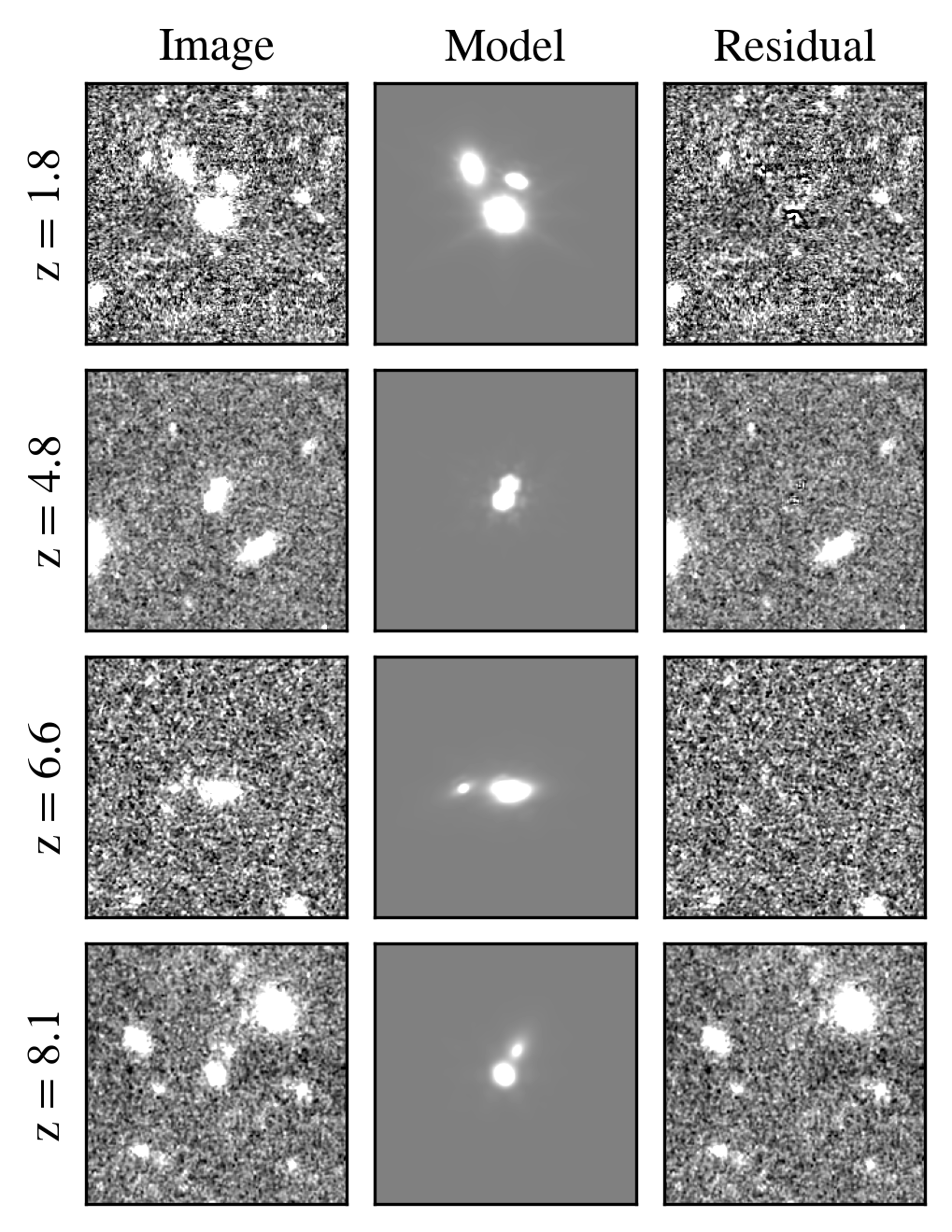}
    \caption{Example fits showing the data image, model image, and the residual image (data - model). Each cutout is 6'' $\times$ 6''.  The redshift of each galaxy is shown to the left of the images.}
    \label{fig:example_fit}
\end{figure}

Although the central position of the source is constrained, all other parameters are allowed to vary freely, and a selection process is used to select good fits with physical parameters after fitting is complete, as the overuse of constraints can lead to \galfit\ converging on unphysical results. Example fits can be seen in \autoref{fig:example_fit}.

\subsection{Comparison with \imfit}
\label{subsec:imfit}
\begin{figure}
    \centering
    \begin{subfigure}{\columnwidth}
        \includegraphics[width=\linewidth]{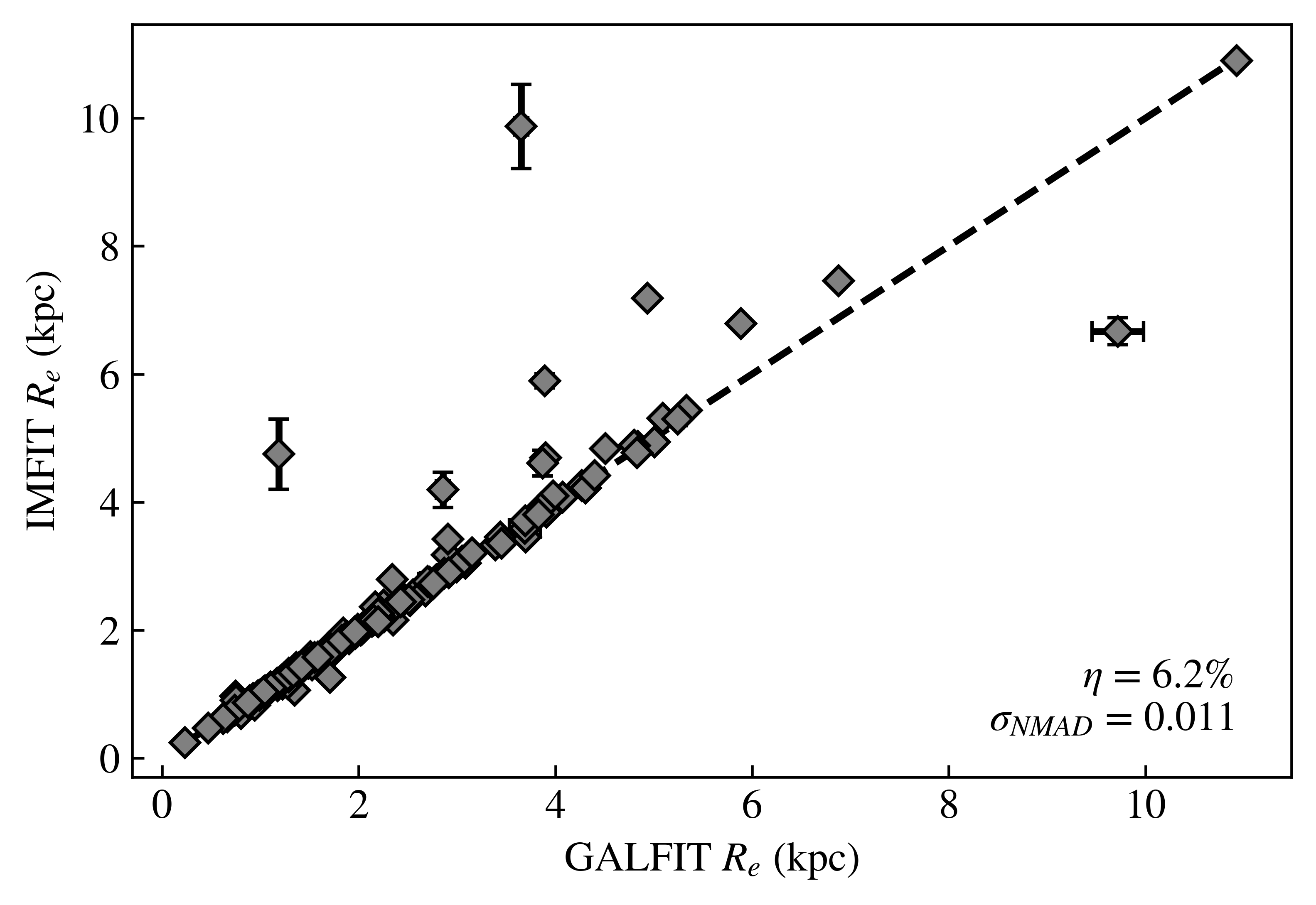}
        \caption{Size comparison.}
        \label{fig:rad_comp}
    \end{subfigure}
    \begin{subfigure}{\columnwidth}
        \includegraphics[width=\linewidth]{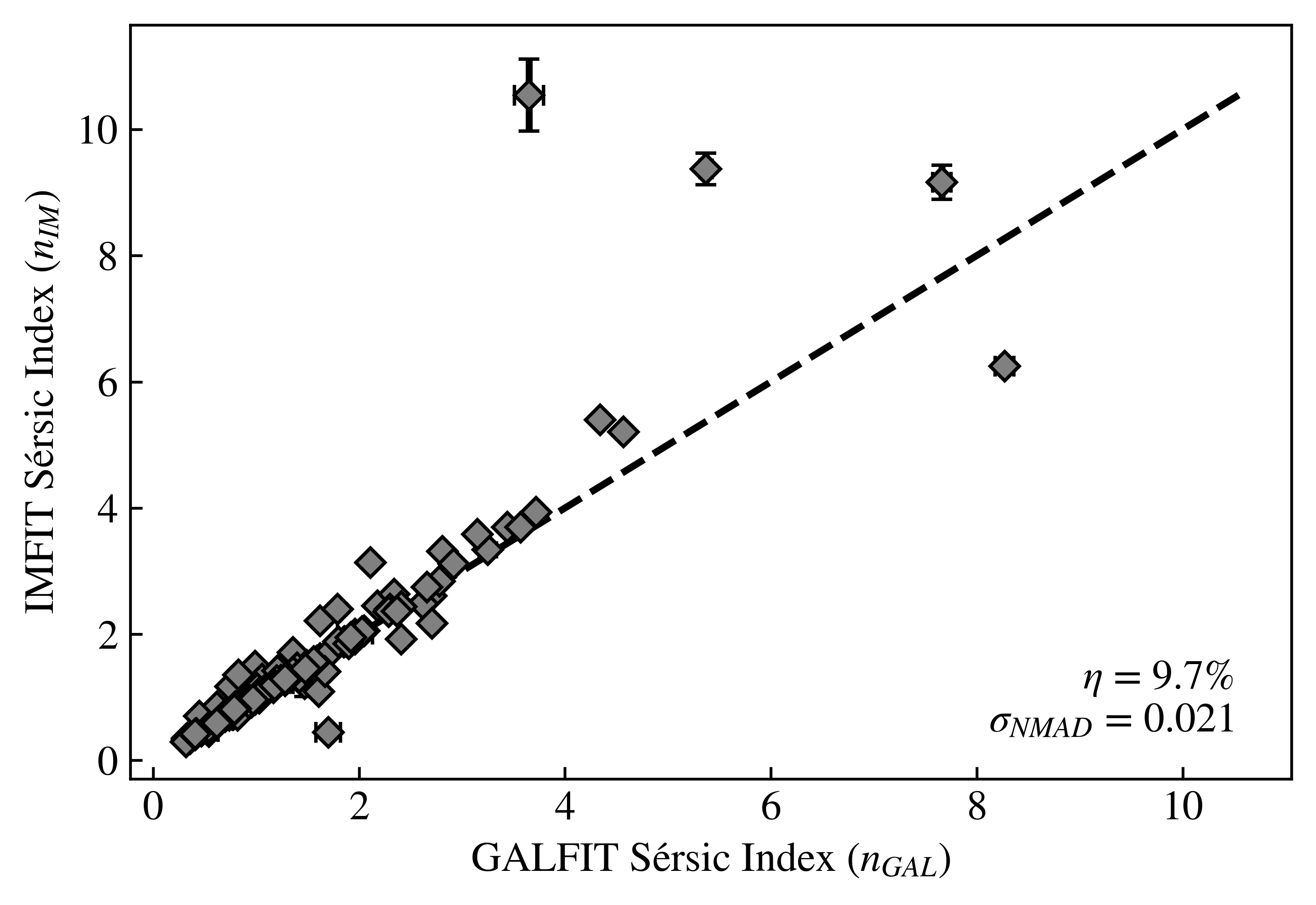}
        \caption{Sérsic index comparison.}
        \label{fig:sersic_comp}
    \end{subfigure}
    \caption{A comparison of two measures - size (top) and Sérsic index (bottom) - as obtained with \galfit\ and \imfit. The black line shows the one-to-one relation. Size and Sérsic index are generally in good agreement, with more variation at larger values, in particular with the Sérsic index. We show the outlier rate $(\eta)$ and NMAD in the bottom right corner of each figure. }
    \label{fig:comparison}
\end{figure}

\imfit\ is an alternative light profile-fitting program which uses Levenberg-Marquardt, Nelder-Mead, and Differential Evolution algorithms to find the best fit parameters \citep{imfit}. In order to test the robustness of our method, we present a comparison of best fit half-light radii and Sérsic indices for a representative sample of 146 objects in CEERS Pointing 1, which have \galfit\ fits that meet the selection criteria explained in Section~\ref{sec:sample_selection} and have a stellar mass of \logm $> 9.5$. We run \imfit\ using the same input parameters and initial guesses as those used for \galfit\, and again use the Levenberg-Marquardt algorithm for consistency. 
The results of this comparison are shown in \autoref{fig:rad_comp} and \autoref{fig:sersic_comp} which plot the half-light radii and Sérsic indices measured for these 146 objects using \galfit\ and \imfit, showing a good agreement between the measured values. 

We use two numerical values to provide a further indicator of reliability, and measure these for the sample of 146 objects used across the comparison. Firstly, we use the outlier rate, defined as the fraction of radii/Sérsic indices obtained by \imfit\ that disagrees with the radii/Sérsic indices obtained by \galfit\ by more than 15\% in $(1+x)$, where $x$ is the measured quantity. Secondly, we use the Normalised Median Absolute Deviation (NMAD) \citep{nmad}, which is defined as 1.48 $\times$ median[$|\Delta x| /(1+x)$], where $\Delta x = x_{galfit} - x_{imfit}$. The NMAD is a measure of the spread of the \imfit\ measurements around the \galfit\ measurements, maintaining its reliability when outliers are present. 

For the half-light radius we find an outlier rate of 6.2\% and NMAD of 0.012, and for Sérsic index we find an outlier rate of 9.7\% and NMAD of 0.021, showing a slightly better agreement between codes for half-light radius than Sérsic index. We also find a better agreement with Sérsic index at lower values of Sérsic index.  We see greater disagreements for a few objects at higher Sérsic index $n$ due to larger contrast which exists at higher values of $n$.  A slight change in the fitting will provide a larger change in $n$ when $n$ is larger.  Overall, however, we find a good agreement between these different codes and use the \galfit\, results throughout the rest of this paper.

\section{Sample Selection}
\label{sec:sample_selection}
We select massive galaxies with stellar masses \logm $>$ 9.5, and then make further selections based on the goodness of fit achieved by \galfit\ for each object. We do not make selections based on the $\chi^{2}_{\nu}$ obtained, which is only used by \galfit\ to determine when it has reached the best fit. The $\chi^{2}_{\nu}$ is provided for the fit as a whole, so in cases where multiple objects are modelled simultaneously, this is not a measure of the goodness-of-fit for only the target object. Instead, we make our own selections based on the output parameters and the residual flux fraction (RFF), which we discuss in the following sections.

\subsection{\galfit\ Parameters}
\label{sec:galfit_params}
In order to remove extreme cases, a fit must meet all of the following criteria:
\begin{enumerate}
    \item The half-light radius must be within 0.01 $< R_e \mathrm{ (pixels)} < $ 100. This ensures that fits where \galfit\ has reached the minimum size possible or where the model would be larger than the cutout size are excluded. 
    \item  The fit Sérsic index lies within the range 0.05 $< n <$ 10.
    \item The fit axis ratio must be $(b/a)$ $>$ 0.01, removing unphysical models. This is particularly prevalent in faint sources, where \galfit\ sometimes converges upon a `bad' fit with a small axis ratio \citep{vanderwal2012}. We do not make selections based upon \texttt{GALFIT} magnitude, and all models are fit by \galfit\ with a rest-frame optical magnitude $<28$.
\end{enumerate}
Where neighbouring objects are being simultaneously modelled, these criteria are only applied to the best fit parameters of the central object. Where \galfit\ does not converge, and gives no best fit parameters, or the best fit parameters do not meet the above criteria, we reject the fit.  We start the fitting process with a sample of 1649 galaxies with \logm $> 9.5$, and through our selection criteria, we reject 192 galaxies, for which we repeat the fitting process with Sérsic index held at a value of $n = 1$. The `Fixed Sérsic' fits must then meet the above criteria for all other parameters, and out of these 192 galaxies, we still reject 93 galaxy fits, thus our sample contains 99 galaxies with a Sérsic index fixed at $n=1$, and 1457 objects with a free value of $n$ at this stage, with a total sample size of 1556 galaxies.

\subsection{Residual Flux Fraction}
We calculate the residual flux fraction for our fits that met the previous criteria in \autoref{sec:galfit_params}. The residual flux fraction (RFF) is a measure of the signal in the residual image that cannot be explained by background fluctuations \citep{hoyos2012}. As in \citet{margalef-bentabol2016}, we define this as 

\begin{equation}
\mathrm{RFF} = \frac{\displaystyle\sum_{(\mathrm{j}, \mathrm{k}) \in \mathrm{A}}\left|\mathrm{I}_{\mathrm{j}, \mathrm{k}} - \mathrm{I}_{\mathrm{j}, \mathrm{k}}^{\mathrm{GALFIT}}\right| - 0.8 \displaystyle\sum_{(j, k) \in A} \sigma_{\mathrm{B}} \mathrm{j}, \mathrm{k}}{\mathrm{ FLUX\_AUTO }}
\end{equation}

\noindent where ${\rm I}$ is the NIRCam image of the galaxy, ${\rm I}^{\textrm{GALFIT}}$ is the model image created by \galfit, $\sigma_B$ is the background RMS image, and FLUX\_AUTO is the flux of the galaxy calculated by \sextractor, all of which are in the rest-frame optical filter of the object. The factor of $0.8$ in the numerator ensures that the expected value of the RFF is 0 for a Gaussian noise error image \citep{hoyos2011}. We calculate the RFF within the Kron radius of the galaxy, where we define the Kron radius as the semi-major axis of the Kron ellipse. Calculating the RFF over a large radius leads to the RFF decaying to zero, where the outer areas can dominate the calculation, even if there is a complex residual at the centre of the image. 

We calculate the background term following the method used in \citet{margalef-bentabol2016}, where we assume that :
\begin{equation}
\sum_{(\mathrm{j}, \mathrm{k}) \in \mathrm{A}} \sigma_{\mathrm{B} j, \mathrm{k}}=\mathbf{N}\left\langle\sigma_{\mathrm{B}}\right\rangle,
\end{equation}
where $\left\langle\sigma_B\right\rangle$ is the mean value of the background sigma for the whole image. We calculate this by placing apertures on blank areas of sky in the image `ERR' extension that we use for creating the \galfit\ sigma images (see \autoref{subsec:pipeline}), and calculating the mean value of these regions. The value of $N$ is the number of pixels within the radius that we are using for the RFF calculation.

In order to remove any remaining objects that are poorly fit with large residuals, we complete visual checks to select an appropriate RFF cutoff value. As a result, we select objects with an RFF value below 0.5. This enables us to remove objects where the light is either very over- or under-accounted for, yet also allows for features that are not modelled precisely (due to features such as spiral arms and bars not being accounted for in single Sérsic fits) but where the measured properties are otherwise reasonable. Through RFF measurements, we reject a further 161 fits due to large residuals.  This results in a final sample of 1395 robust galaxy fits, of which 1313 (94.1\%) were fit with a free value of $n$, and 82 (5.9\%) were fit with a fixed value of $n=1$, which we further analyse in \autoref{sec:results}. The rejected fits are mostly comprised of lower mass galaxies, although there are fits rejected at all masses within our sample, and we do not see a change in RFF as a function of redshift. 

\subsection{Final Sample}
We begin the fitting process with a sample of 1649 galaxies, and after our quality cuts we recover 1395 galaxies for our final sample. This recovery rate of 84.6\% is higher than comparable analyses, such as \citet{Suess2022} ($\sim 60\%)$, although we repeat the fitting process with a fixed Sérsic index where the first fitting procedure has failed. However, fits with a fixed Sérsic index only account for 5.9\% of our sample. We note that we have not made any selections based upon the magnitude of the \galfit\ model, as the input and output magnitudes are in good agreement.  In the rest-frame optical, $3.66\%$ of the \galfit\ best-fitting models have a half-light radius in pixels that is smaller than the FWHM of the PSF, thus almost all of our sources are resolved. 
This sample is used throughout this paper to carry out our analyses of the evolution of galaxy size and structure.

\section{Results}
\label{sec:results}

In the following sections we describe the results of our analysis.  This includes examining the sizes of our systems, as well as the overall shapes of these galaxies based on their Sérsic indices and how these evolve with time. Where we state results for passive and star forming galaxies, whereby these are defined by having a specific star formation rate (sSFR) below or above the midpoint of the distribution of values within the redshift bin. We calculate the specific star formation rate using the star formation rates and stellar masses from \citet{duncan2014, duncan2019} (see \autoref{subsec:redshift} for full details). We then define a galaxy with a specific star formation rate greater (lower) than the median sSFR within the redshift bin, to be a star-forming (quiescent) galaxy. While we use the half light radius as a measure of size throughout this paper, it is worth noting that when we compared our trends to those obtained with another size measure, such as $R_{90}$, our findings do not change. The use of $R_e$ also allows for a continuation of work completed with HST, and a wide comparison of the following results to previous work in this area.

\subsection{Half-light radii and Sérsic indices}
\begin{figure*}
    \centering
    \includegraphics[width=0.8\textwidth]{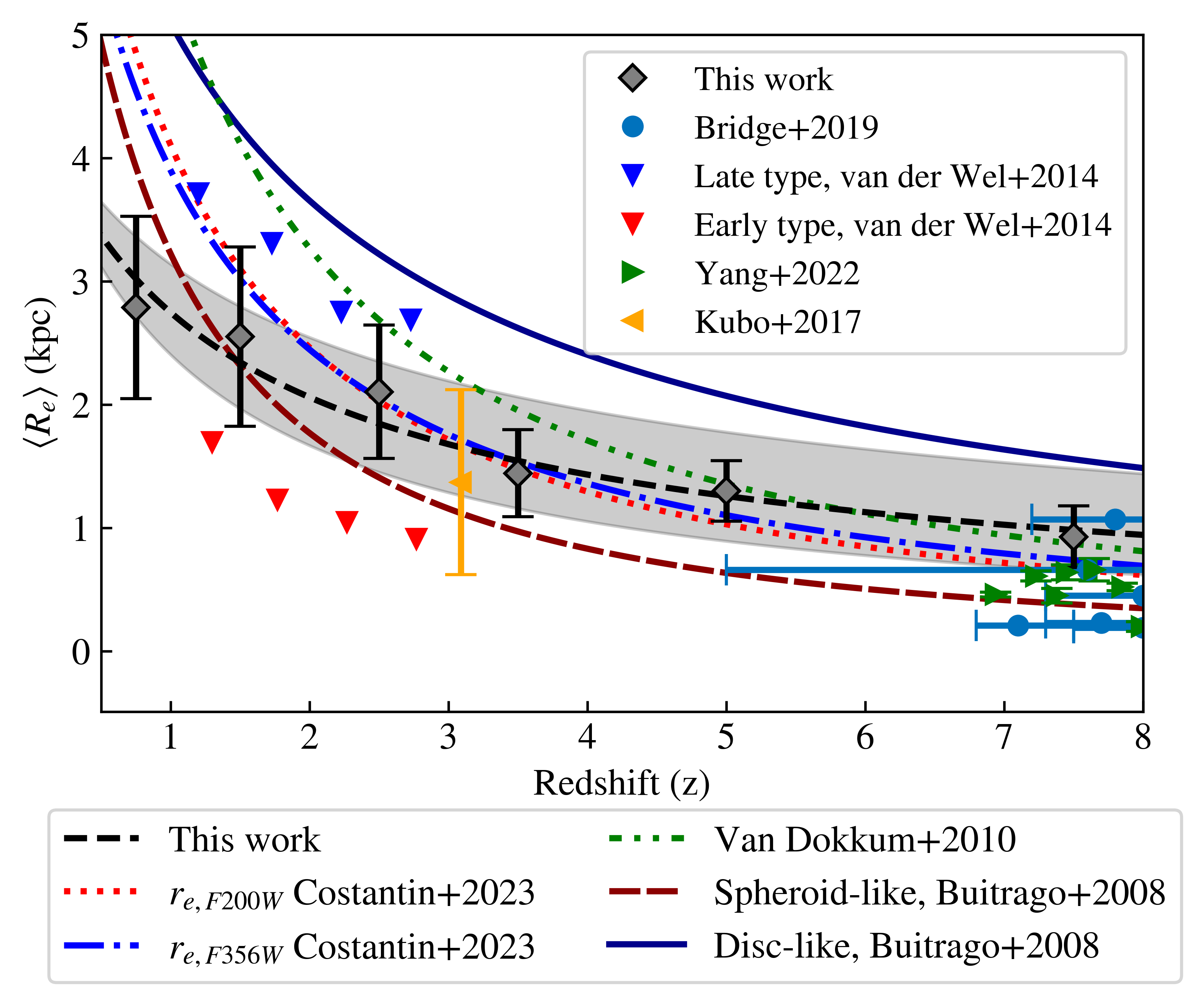}
    \caption{Power law fit showing the size evolution of all galaxies within our sample, compared to results from previous work. The black dashed line is of the form $R_e (kpc) = 4.50\pm1.32(1+z)^{-0.71\pm0.19}$,  with the grey, shaded area showing the error on the power-law fit. The grey diamond points are the mean galaxy sizes in each redshift bin, and error bars are $1\sigma$ in length. The previous work shown in the Figure uses HST data, except for \citet[][]{Yang2022}, which uses JWST data, \citet{dokkum_2010} which uses NOAO/YaleNEWFIRM Medium Band Survey Data, and \citet{costantin2023} which uses the TNG50 simulation to produce mock CEERS observations. We note that we only plot redshift errors for \citet{bridge_2019}, as radius errors are not provided. }
    \label{fig:rad_evolution}
\end{figure*}

We measure the half-light radii of our objects using \galfit, and convert the values from pixels to their physical half-light radii in kpc. \autoref{fig:rad_evolution} shows the size evolution with redshift for our sample of 1395 galaxies. We use the radius from the filter nearest to the rest-frame optical for each object, and find that the sizes are well fit by the power-law relation:

\begin{equation}
\langle{\rm R_{e}}\rangle = 4.50 \pm 1.32 (1+z) ^{-0.71 \pm 0.19}.
\end{equation}

\noindent This is such that the average sizes of our sample, in terms of effective radii ($\langle{\rm R_{e}}\rangle)$, become progressively smaller at increasing redshifts.    This trend for galaxies at a given mass selection to become smaller at higher redshifts had been known to exist at $z < 3$ for many years \citep[e.g.,][]{Trujillo2007, Buitrago2008,vanderwal2012}, yet this is the first time this has been shown using JWST observations for similar types of studies. We also compare our power law function to those derived in comparable studies, shown in \autoref{fig:rad_evolution}. We note that the curves presented in \citet{Buitrago2008} are normalised with respect to SDSS data, thus we perform an arbitrary normalisation of these curves to align them with the scale employed.  We also extrapolate all curves to cover our entire redshift range. We compare to a range of individual points and power-law curves. 

It is important to note that the evolution of galaxy size and Sersic index are highly dependent on the redshift ranges studied and the stellar mass and/or magnitude ranges which are included in the analysis.  For example, if we study just the super massive galaxies at \logm $ > 11$, we would observe that systems have a stronger evolution than at lower masses \citep[e.g.,][]{Buitrago2008, Buitrago2013}.  As such it is important to establish a clear basis for comparison, as no previous study has measured galaxies using the exact same approach as in this study. The points from \citet{vdw2014} are for galaxies with \logm $\sim 10.75$, \citet{bridge_2019} and \citet{kubo_2017} are at \logm $\sim 10$, and \citet{Yang2022} select bright objects based on their magnitudes in the F444W band. The curves given are for a mass selection of \logm $> 9$ at $3 \leq z \leq 6$,\citep{costantin2023}, \logm $ > 11$ at $1.7 < z < 4$ \citep{Buitrago2008}, and for a number density based selection at $z < 2$\citep{dokkum_2010}. We extrapolate these curves to cover our entire redshift range.

 \autoref{fig:rad_evolution}  shows that there is a steep evolution for galaxies within the mass range employed in this paper. We find, as previously mentioned, that the size evolution of galaxies follows a power law with the shape $\sim (1+z)^{-0.71}$, which is less steep than previous results when comparing the evolution up to $z \sim 3$.  This is partially due to the fact that we are observing galaxies at higher redshifts where the size evolution tapers off and does not continue as steeply at higher-z. It is important to keep in mind that redshift (z) values do not scale linearly with time, and there is much more time at a given $\delta z$ at low redshift than at the higher redshifts.  Another reason for the difference, can be seen at the lowest redshifts, where our galaxies are on average smaller than the previous work. This is likely due to us using a lower mass cut to define our sample of galaxies, resulting in on average smaller systems. This is consistent with findings from simulations, where it has been shown that galaxy size correlates with stellar mass, thus resulting in lower mass samples having smaller sizes, on average \citep{furlong_2017, ma_2018}.

We also investigate the difference in the size-redshift relation for populations of galaxies with high and low Sérsic indices, defined as $n>2$ and $n<2$ respectively, although using $n = 2.5$ to separate the sample produces effectively the same results.   We do this to determine how the size evolution depends on the shape of the profile, with galaxies at $n > 2$ possibly more like the massive galaxies we see in the local universe and those with $n < 2$ possibly progenitors of disc galaxies or those undergoing mergers.
 
\begin{figure}
    \centering
    \includegraphics[width=\columnwidth]{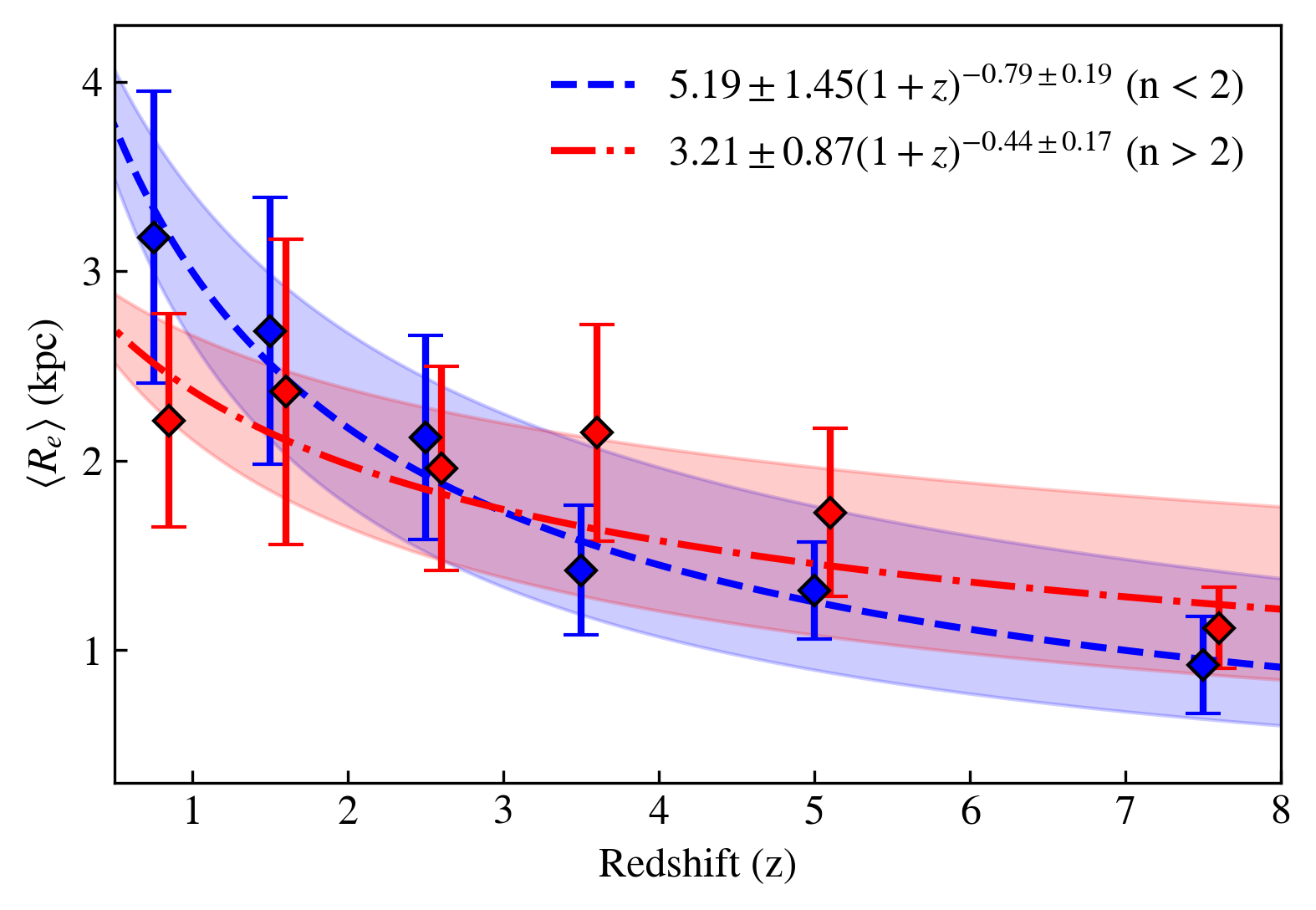}
    \caption{Power law fit for galaxies separated into two groups by Sérsic index as defined at $n = 2$, where $n <2$ represents disc-like galaxies, and $n > 2$ represents elliptical-like galaxies. The sizes of these objects mostly diverge at the lowest redshifts. The  diamond points are the mean galaxy sizes in each redshift bin, and the error bars are $1\sigma$ in length. The shaded region around each line represents the error on the power-law fit.}
    \label{fig:radius_sersic}
\end{figure}

 As can be seen in \autoref{fig:radius_sersic}, at low redshift, the galaxy populations have a clearly different size-redshift relation, but at redshifts higher than $z \sim 3$, the relations show a greater similarity, suggesting that effective radius is less dependent upon the Sérsic index at high redshift compared to lower redshifts.  What this means is that galaxies do not differentiate between overall morphology, as measured by the Sérsic index, until around $z \sim 3$, consistent with findings at $z < 5$, where star forming galaxies exhibit inside out growth \citep{roper_2023}. This suggests that this aspect of the `Hubble Sequence' was in place by at least $z \sim 3$, with a disc-like ($n < 2$) and elliptical-like ($n > 2$) population clearly defined. 

As shown in \autoref{fig:radius_sersic} and \autoref{fig:rad_evolution}, we also find that galaxies at higher redshift are almost all compact objects, regardless of their Sérsic index, and are smaller than their low-redshift counterparts of similar mass, in agreement with previous studies and simulations \citep[e.g,][]{Trujillo2007, Buitrago2008,vdw2014, costantin2023}.  Furthermore, \autoref{fig:radius_sersic} shows that the sizes of the high and low Sérsic index populations evolve differently at $z < 3$, with the objects with lower Sérsic indices following a steeper size - redshift relation, suggesting a difference in structural growth mechanisms in each population, likely due to the onset of inside out growth in these systems with lower Sérsic indices. Overall, we also show that the compact systems of the early universe are not representative of the galaxy population today, suggesting a strong size evolution must occur, continuing until the present day.  This growth is such that we find an increase of a factor of three from $z \sim 7$ to $z\sim 1$, with roughly a doubling of size from $z \sim 7$ to $z \sim 3$, a time period of  $\sim 1.4$ Gyr.  The relation is also still evolving at the highest redshifts, confirming that galaxy evolution as well as evolution in structure were already taking place in the first Gyr since the Big Bang. 

\begin{figure}
    \centering
    \begin{subfigure}{\columnwidth}
        \includegraphics[width=\columnwidth]{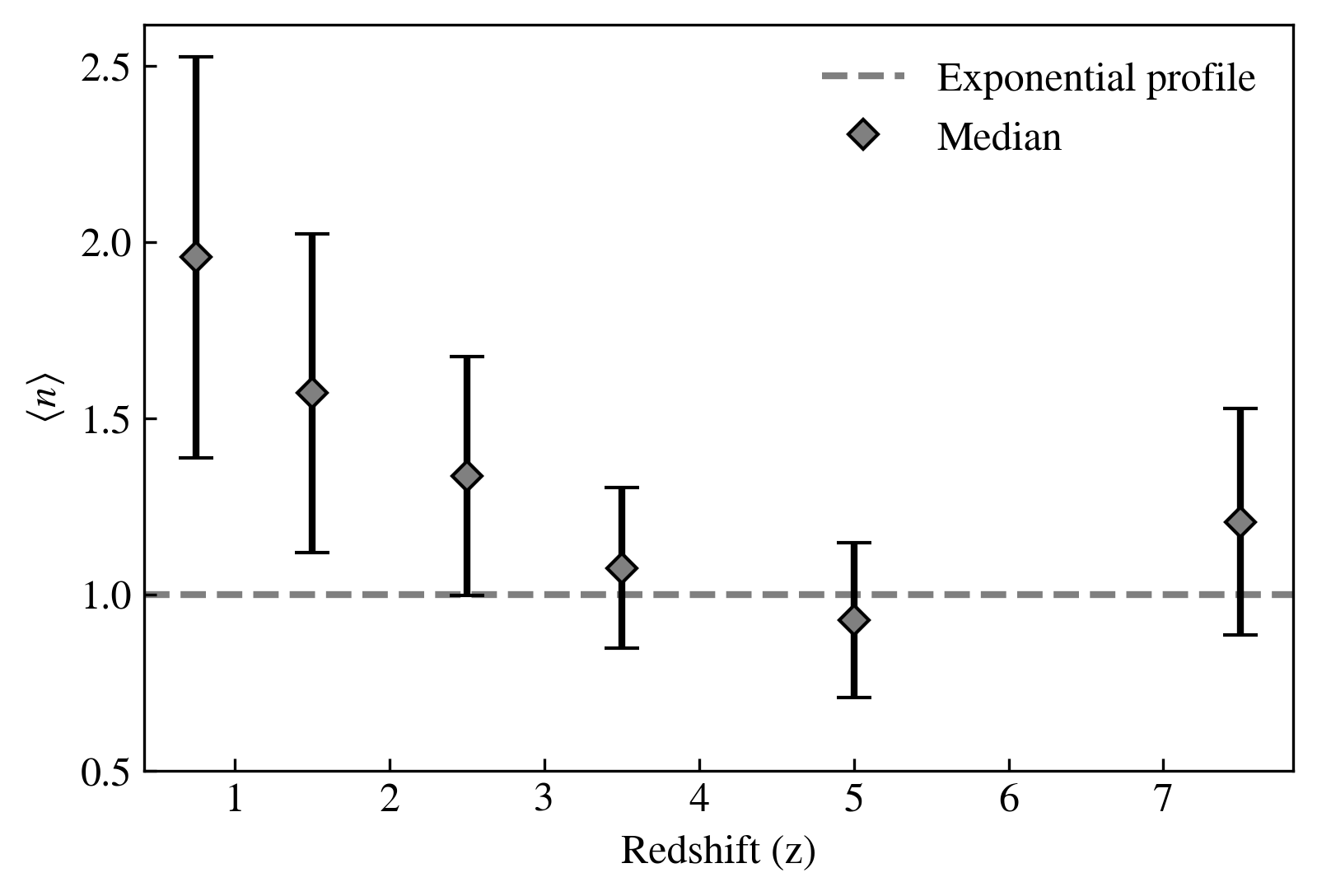}
        \caption{Sérsic - redshift distribution for all galaxies within our sample.}
        \label{fig:sersic_all}
    \end{subfigure}

    \begin{subfigure}{\columnwidth}
        \includegraphics[width=\columnwidth]{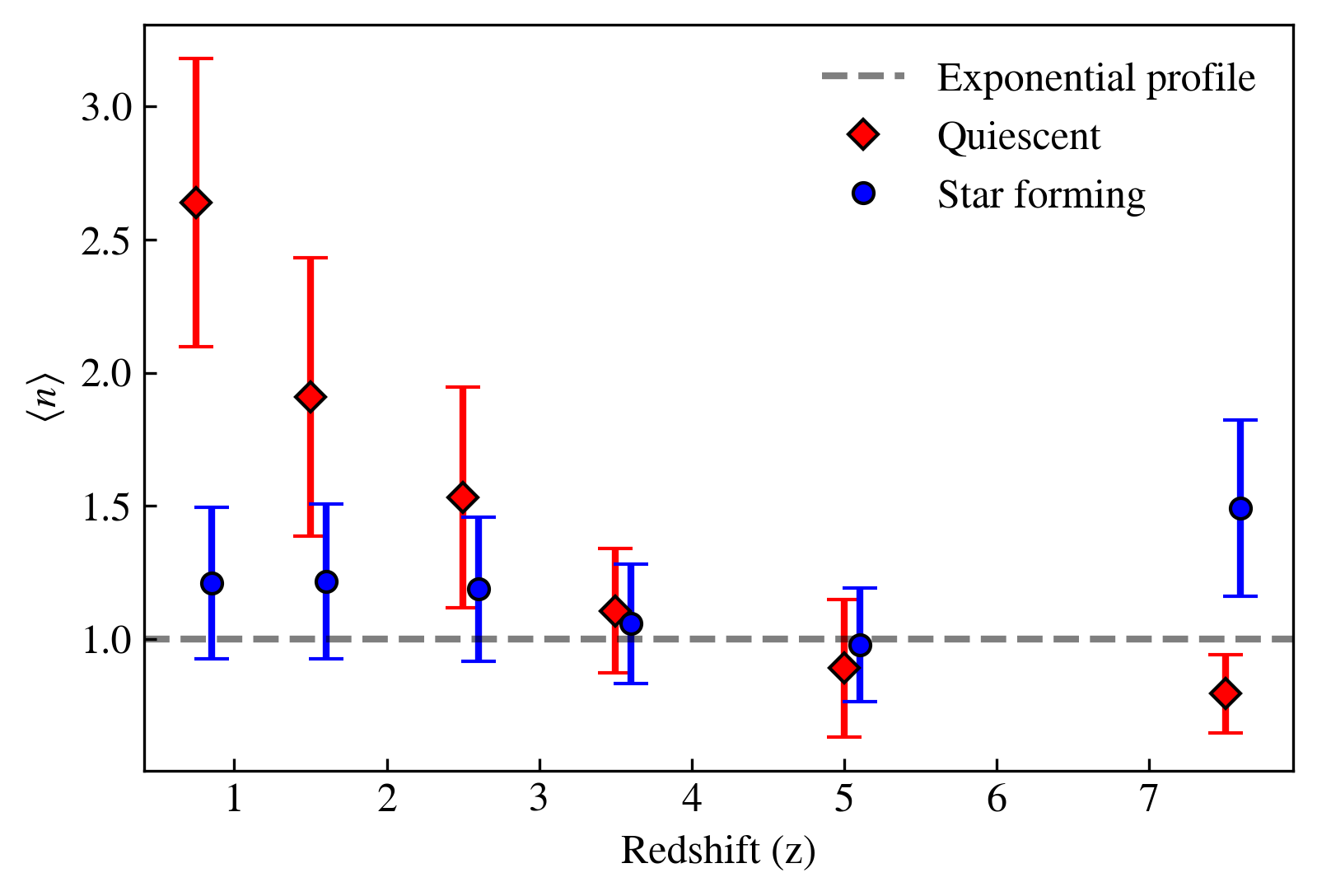}
        \caption{Sérsic - redshift distribution for passive and star forming galaxies.}
        \label{fig:sersic_ssfr}
    \end{subfigure}
    \caption{The Sérsic index - redshift relations for all galaxies (top), and passive and star forming populations, relative to the average sSFR in each redshift bin (bottom). Large diamond and circle points are median values of each redshift bin, with error bars $1\sigma$ in length. The grey dashed line at $n = 1$ represents the special case of the exponential disc profile.}
    \label{fig:sersic_vs_z}
\end{figure}

For those objects fit with a free Sérsic index, we  also investigate the evolution of Sérsic index with redshift, as shown in \autoref{fig:sersic_vs_z}. \autoref{fig:sersic_all} shows that for all free-fit $n$ galaxies within our sample, Sérsic index decreases with increasing redshift, suggesting a higher proportion of disc-type galaxies in the early universe.   This is in agreement with findings shown in \citet{robertson_2023}, where it is shown disc candidates appear at high redshifts of $z\sim5$ with $\langle n \rangle = 1.04$.   Whilst these objects have Sérsic indices similar to modern pure disc galaxies, this does not necessarily imply that these are rotating disks \citep[e.g.,][]{buitrago2014}.  This is in contrast to previous findings using HST that there were fewer disc galaxies at $z > 1.5$, although the lower resolution of HST lead to misclassification of galaxies \citep{Ferreira22b}, with the improved depth and quality of JWST imaging when compared to HST imaging revealing disc morphologies previously obscured due to noise \citep{robertson_2023}. Exploring this complicated subject is beyond the scope of this paper.  The slight increase in Sérsic index at $z \sim 7.5$ could be due to an increase of spheroid galaxies reported in \citet{Ferreira22b}, particularly in the star forming population (see \autoref{fig:sersic_ssfr}), as spheroid galaxies have been found to account for a higher proportion of the sSFR budget \citep{ferreira22a} at high redshifts, but could also simply be due to random chance and increased errors at higher redshifts.  We further discuss size and Sérsic index changes as a function of morphology in \autoref{sec:morph}.

In \autoref{fig:sersic_ssfr}, we investigate Sérsic indices for passive and star forming populations, and find that the star forming galaxies have Sérsic indices around $n \sim 1$, suggesting that most star formation takes place within disc galaxies, which is also the case when identifying these systems through visual means \citep[][]{Ferreira22b}. This is also in agreement with measurements of star forming galaxies, which are found to be typically discy with $n~\sim 1.2$ \citep{paulino-afonso}. The slight deviation from this trend at the highest redshifts could be due to stars forming in young, compact sources, before evolving into disc-like galaxies. The populations start to show differing trends at $z \sim 3$, again hinting at the establishment of a bifurcation of the galaxy population in structure around this time. This shows that the star forming and passive galaxy populations possibly evolve through different mechanisms to create differing physical properties at later times, such as inside out star formation in star forming galaxies, and stellar migration in passive systems.

\subsection{Galaxy Size-Mass Distribution}
\label{sec:size-mass}
\begin{figure*}
    \centering
    \includegraphics[width = 0.6\textwidth]{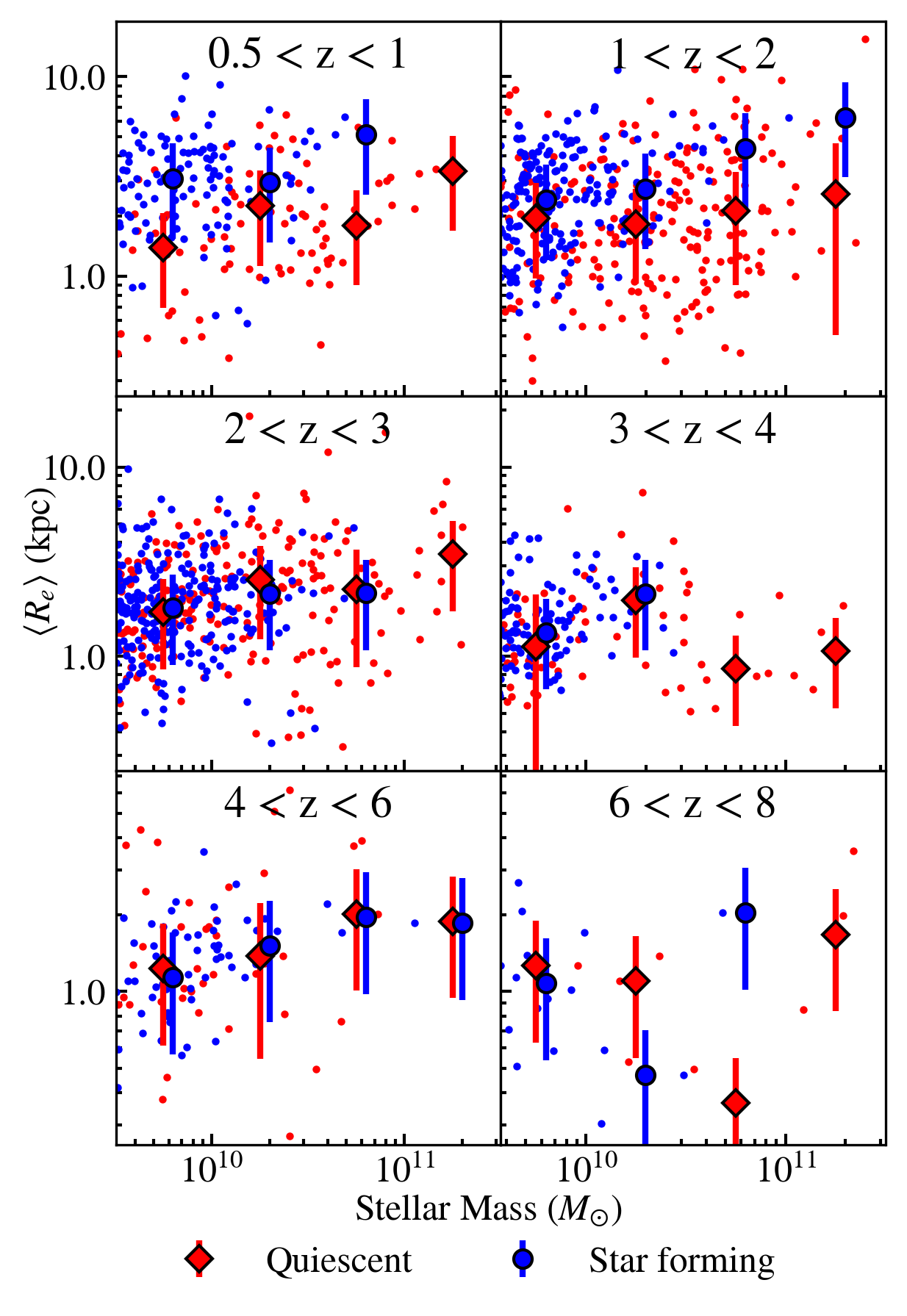}
    \caption{The size-stellar mass distribution of `passive' and `star forming' galaxies, relative to the midpoint of the specific star formation rate in each bin, as explained in \autoref{sec:size-mass}. The larger points represent the median values in mass bins. We use a 50\% error floor, with error bars one standard deviation in length, or representing a 50\% error on the median, in cases where the error floor is larger.}
    \label{fig:mass}
\end{figure*}

We plot the galaxy size-mass relation in different redshift bins, as shown in \autoref{fig:mass}. For each redshift bin, we separate galaxies into either a star-forming or passive population with respect to the average value in the given redshift bin, using the median specific star formation rate (sSFR) of each bin as the separating value.  This midpoint is determined by plotting a histogram of all the sSFR values for galaxies within a given redshift bin. The median value is then measured, and galaxies which are lower than this we call `passive', and those above this `active',relative to the average sSFR in the given redshift bin. This is the same method we used previously for separating star forming from  non-star forming galaxies when investigating trends with size and Sersic index.

In general, we find that galaxy size increases with stellar mass, in both the quiescent and star forming populations up to $z \sim 6$, although this tends to be more obvious for the star-forming galaxies.  We also see that the star forming galaxies at the lower redshifts are nearly always larger at a given mass than the passive galaxies, although this tends to break down at the lower masses. 

We also find that at redshifts higher than $z > 3$, the sizes of the quiescent and star forming populations are statistically the same, suggesting that at a fixed mass at high redshift, quiescent galaxies may not be smaller than star-forming galaxies, as predicted in \citet{Suess2022}, likely due to transient quiescence driven by high redshift active galactic nuclei (AGN) activity \citep{lovell_2023}.   This means that whatever is differentiating the sizes of galaxies as a function of shape or star formation rate does not come into play until past redshift $z \sim 3$.  This implies that the physical processes of growing galaxies is truncated for the passive galaxies, even if these systems still acquire stellar mass throughout their history.  We discuss possible reasons for this later in the discussion section.

\subsection{Changes in Effective Radii as a Function of Observed Wavelength}

The changes in a galaxy's appearance and size as a function of wavelength can provide many clues to the formation history and stellar populations of modern galaxies \citep[e.g.,][]{Windhorst2002,TaylorMager2007, Mager2018, Suess2022}.  An example of this is that if a galaxy forms inside-out, with the older stellar populations in the centre of the galaxy, then most likely these systems would appear larger at shorter wavelengths, and vice-versa if formation occurred via outside-in. 

\begin{figure*}
    \centering
    \includegraphics[width = 0.8\textwidth]{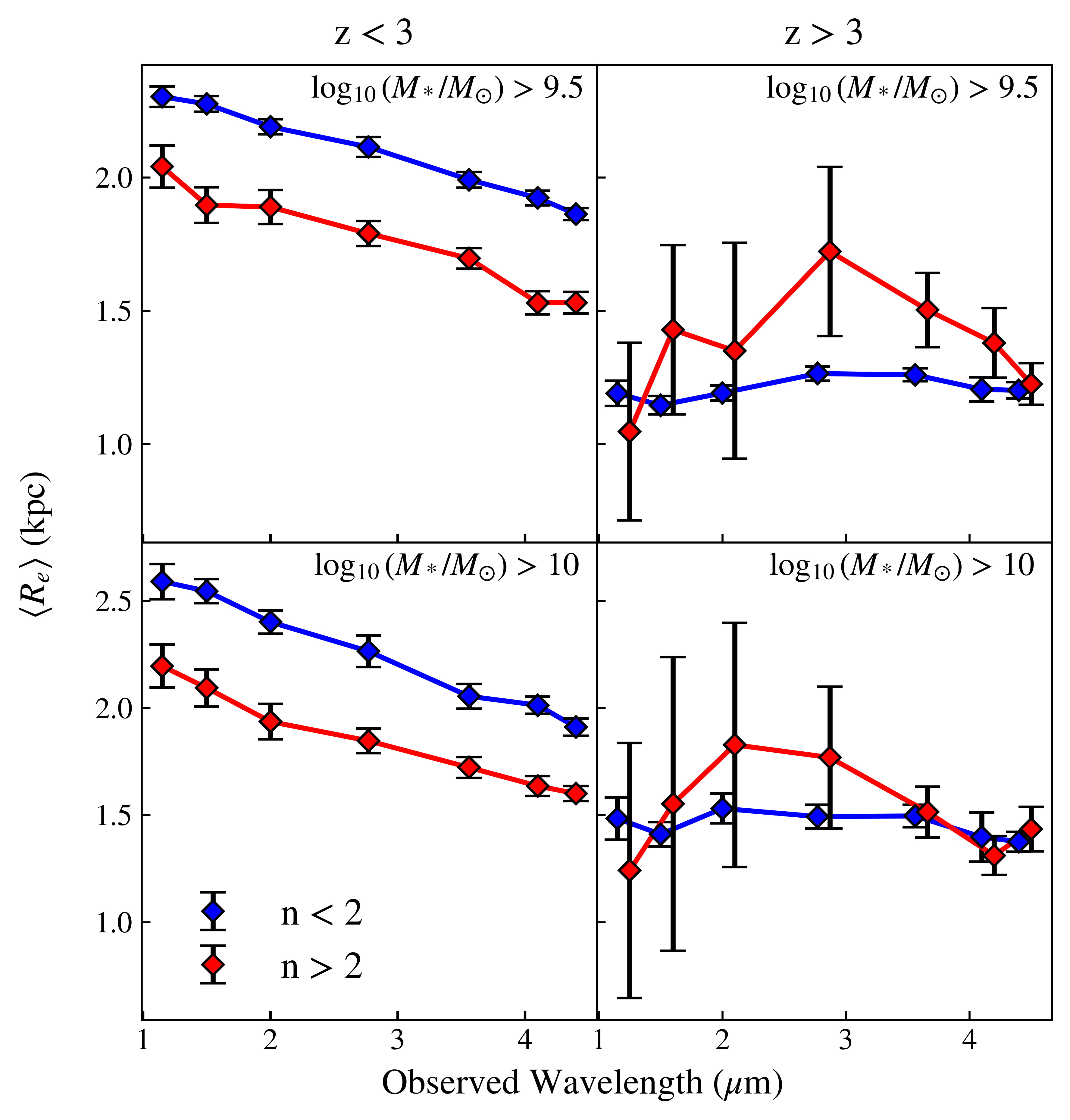}
    \caption{The half-light radius as a function of wavelength. These plots show the radius evolution for $z < 3$ (left panels), and $z > 3$ (right panels) galaxies. In each redshift bin, we show the evolution for low-mass and high-mass galaxies, with the galaxies separated into high-Sérsic (red) and low-Sérsic (blue) index populations.  These figures show that up to $z \sim 3$ (left panels), galaxies are more compact when measured in redder filters, regardless of selection method. Past $z=3$ (right panels), the relation is flatter, with the galaxies having $n < 2$ being smaller at this stage.}
    \label{fig:rad_wavelength}
\end{figure*}

As such, we measure half-light radii in all available wavelengths for our sources, and use these sizes to probe the size evolution of our galaxy sample as observed at different wavelengths. As seen in \autoref{fig:rad_wavelength}, average galaxy sizes become increasingly more compact and smaller when observed at longer wavelengths, for objects with both high and low Sérsic indices at $z < 3$.  We also see again, that those galaxies with higher Sersic indices are smaller on average at all wavelengths to those systems with lower indices.   This is an indication that galaxy sizes are larger at shorter wavelengths where bluer and young light is probed, due to the formation mechanisms for these galaxies. This would be such that the outer parts of these systems consist of younger stars, compared to their inner portions made of older stars.  More detailed analysis of the colour gradients and star formation gradients of these galaxies would answer this question. We also note that dust attenuation can increase the observed half-light radius \citep[e.g.,][]{marshall_2022, Roper_2022, popping_2022}, although this would require a more in depth observational analysis.

At $z < 3$, we also find that galaxies with higher Sérsic indices have smaller radii in both mass bins, with this effect being more noticeable in the highest mass bin. However, we do not see the same effect past $z \sim 3$, where we see a much flatter relation, suggesting that galaxies at high redshift are forming stars throughout the entire galaxy, with blue and red light emitted throughout. However, when we divide the higher redshift galaxies into different bins, we obtain a noisier trend, and thus we hesitate to draw any further conclusions from this observed trend. For objects with Sérsic indices of $n > 2$, there is much more scatter within the relation, although there are larger errors, due to the smaller number of galaxies with high Sérsic index at $z > 3$. Analysing this in smaller bins, such as $\Delta z = 1$, did not yield different conclusions.

\begin{figure}
    \centering
    \includegraphics[width = \columnwidth]{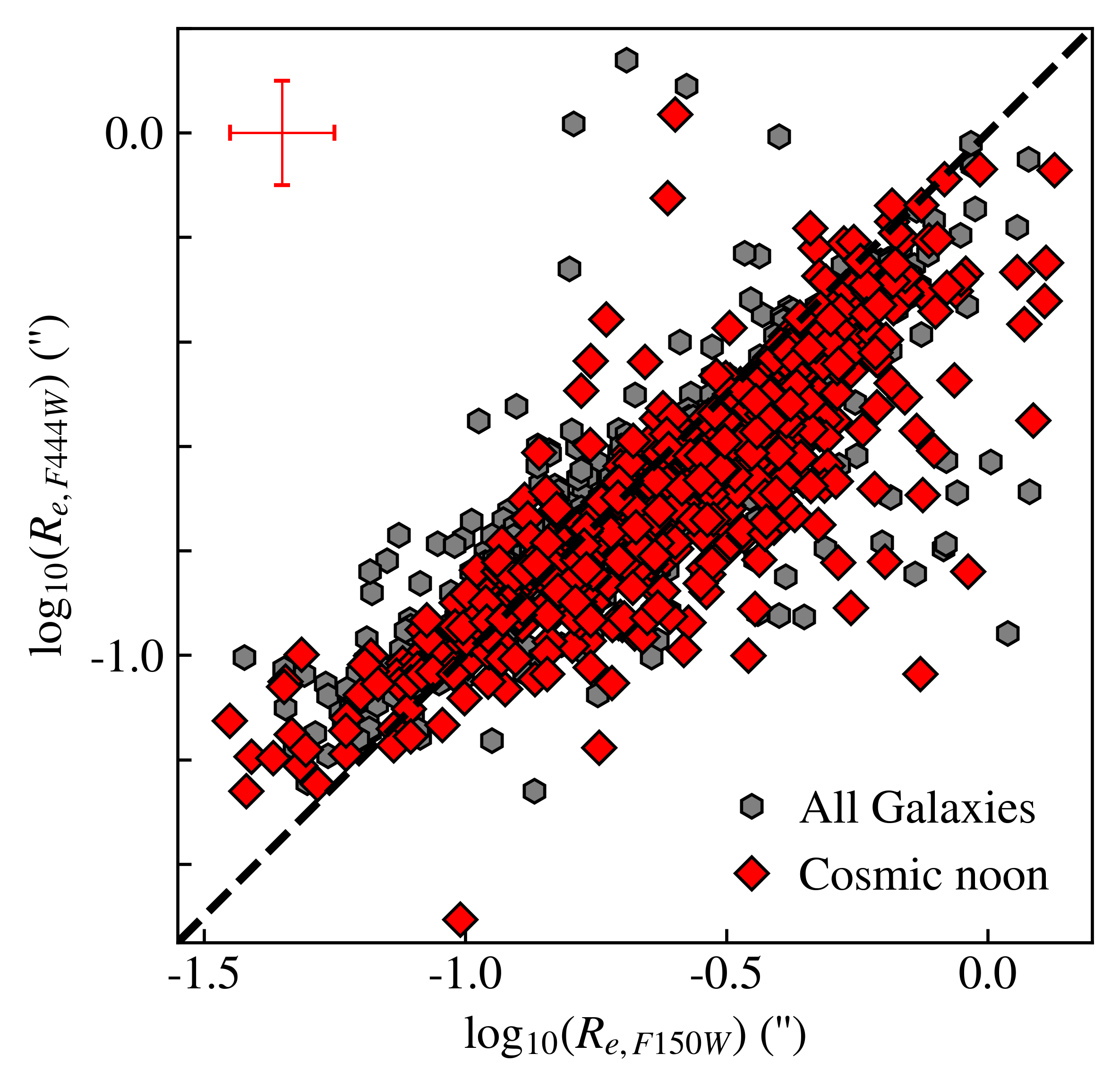}
    \caption{A comparison of sizes measured in the F444W and F150W filters, showing that F444W sizes are smaller than those in F150W. This implies that galaxies measured in the near-infrared with \emph{JWST} are more compact than rest-frame optical sizes previously measured with \emph{HST}.  Galaxies at cosmic noon ($1 \le z \le 2.5$) are shown as red diamonds whilst the grey hexagons represent all galaxies within our sample that do not fall within cosmic noon. These latter galaxies do not show the same effect. The black, dashed line is the one-to-one relation. The error bar represents a typical error of 0.2 dex on these measurements.}
    \label{fig:444_150_comp}
\end{figure}

\begin{figure}
    \centering
    \includegraphics[width = 0.8\columnwidth]{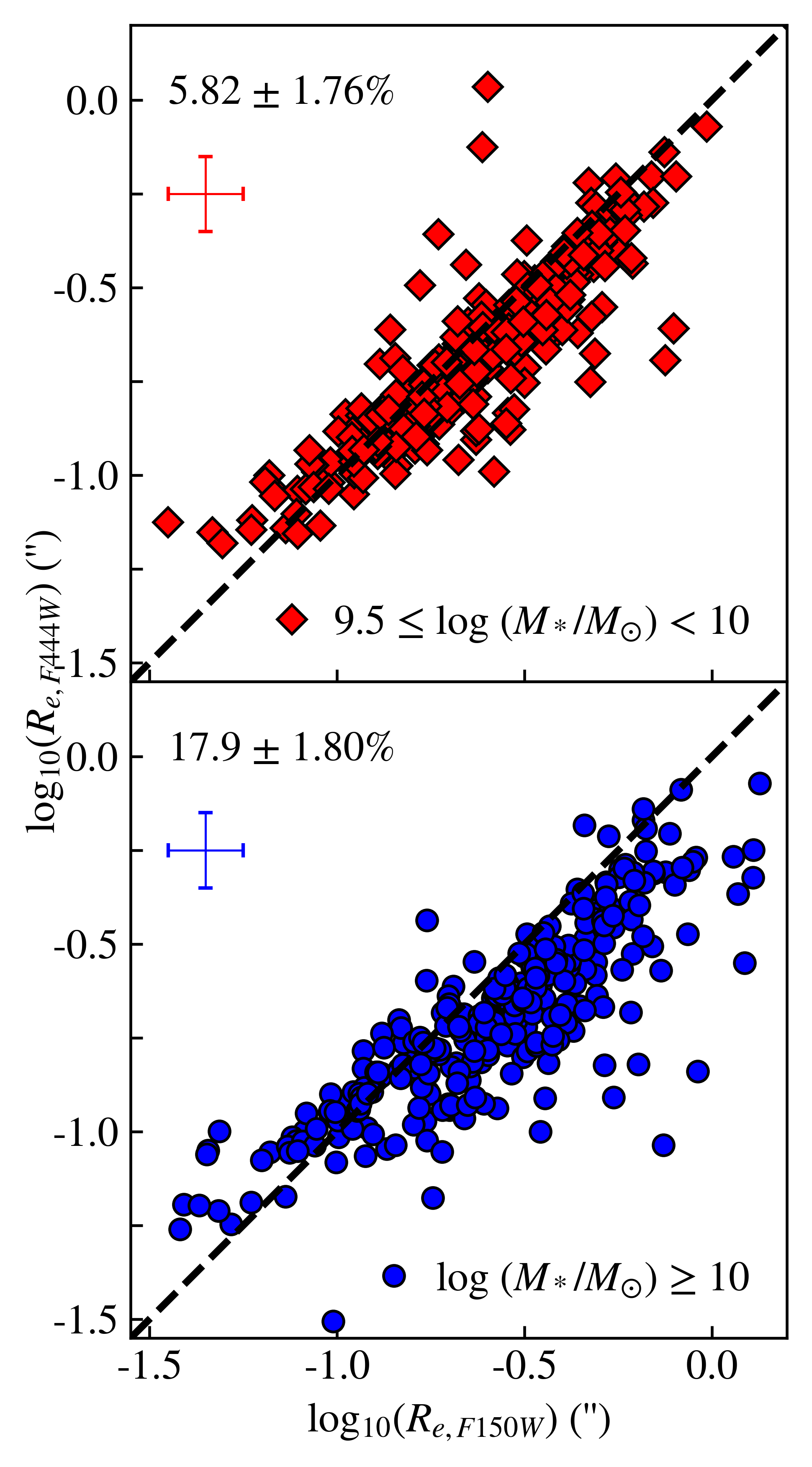}
    \caption{Size comparison for different mass bins at cosmic noon. We find an increasing size difference with increased stellar mass. The black dashed line is the one-to-one relation. The percentages in the upper left corner are the average size difference for each redshift bin. The error bar represents a typical error of 0.2 dex.}
    \label{fig:radius_comp_mass}
\end{figure}

To compare with previous JWST work on similar questions, we also compare the sizes of our galaxies measured in the F444W ($4.4\mu m$) band to the F150W ($1.5\mu m$) band sizes, as shown in \autoref{fig:444_150_comp}. This allows us to probe the rest-frame optical, observed in F150W, and compare with the rest-frame near infrared, observed with F444W. This comparison is quoted in arcseconds, as a comparison of the on-sky sizes. We find that for galaxies at cosmic noon ($1 \le z \le 2.5$), the $4.4\mu m$ sizes are 11.4$\pm$1.28\% smaller than the $1.5 \mu m$ sizes on average, in agreement with the $\sim9\%$ difference found in \citet{Suess2022}. Taking sizes measured in the near-infrared (observed with $4.4\mu m$ at this redshift range) to be a reasonable proxy for stellar mass distributions, this shows that the stellar mass profiles of galaxies are smaller and more compact than their star forming `light' profiles. 
We do not find that the galaxies outside of cosmic noon show the same effect, as the F150W and F444W filters no longer correspond to the rest-frame optical and rest-frame infrared of these galaxies. As we have shown that the smaller appearance in F444W is indeed due to the rest-frame infrared profile being smaller, thus showing the mass profile of the galaxy is smaller than the light profile, we investigate how this varies with the stellar masses of galaxies. 
This effect is dependent on the mass of the galaxy, as shown in \autoref{fig:radius_comp_mass}, where we examine the size difference at cosmic noon in two mass bins. We find that for galaxies with 9.5 $\le$\logm$<$ 10, the F444W sizes are $5.82\pm1.76\%$ smaller than their F150W sizes. This increases with increasing mass, with sizes being $17.9\pm1.80\%$ smaller for objects with 10$\le$\logm.  This implies that on average we see a greater difference in size between different wavelengths for the highest mass galaxies.

\subsection{Correlation with Visual Morphology}
\label{sec:morph}

\begin{figure}
    \centering
    \begin{subfigure}{\columnwidth}
        \centering
        \includegraphics[width=\columnwidth]{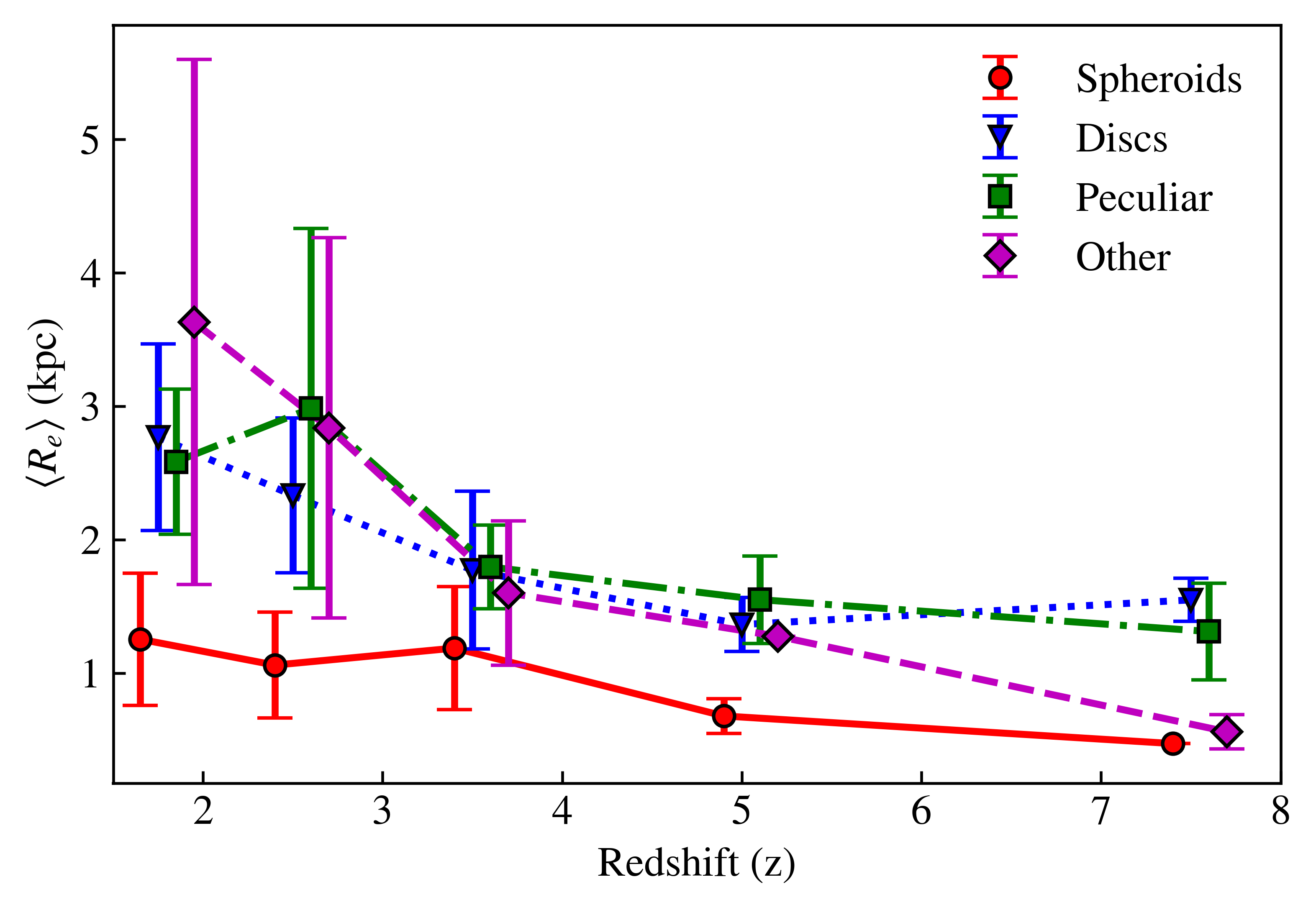}
        \caption{Radius (kpc)}
        \label{subfig:radius_morph}
    \end{subfigure}
    
    \vspace{\baselineskip} 
    
    \begin{subfigure}{\columnwidth}
        \centering
        \includegraphics[width=\columnwidth]{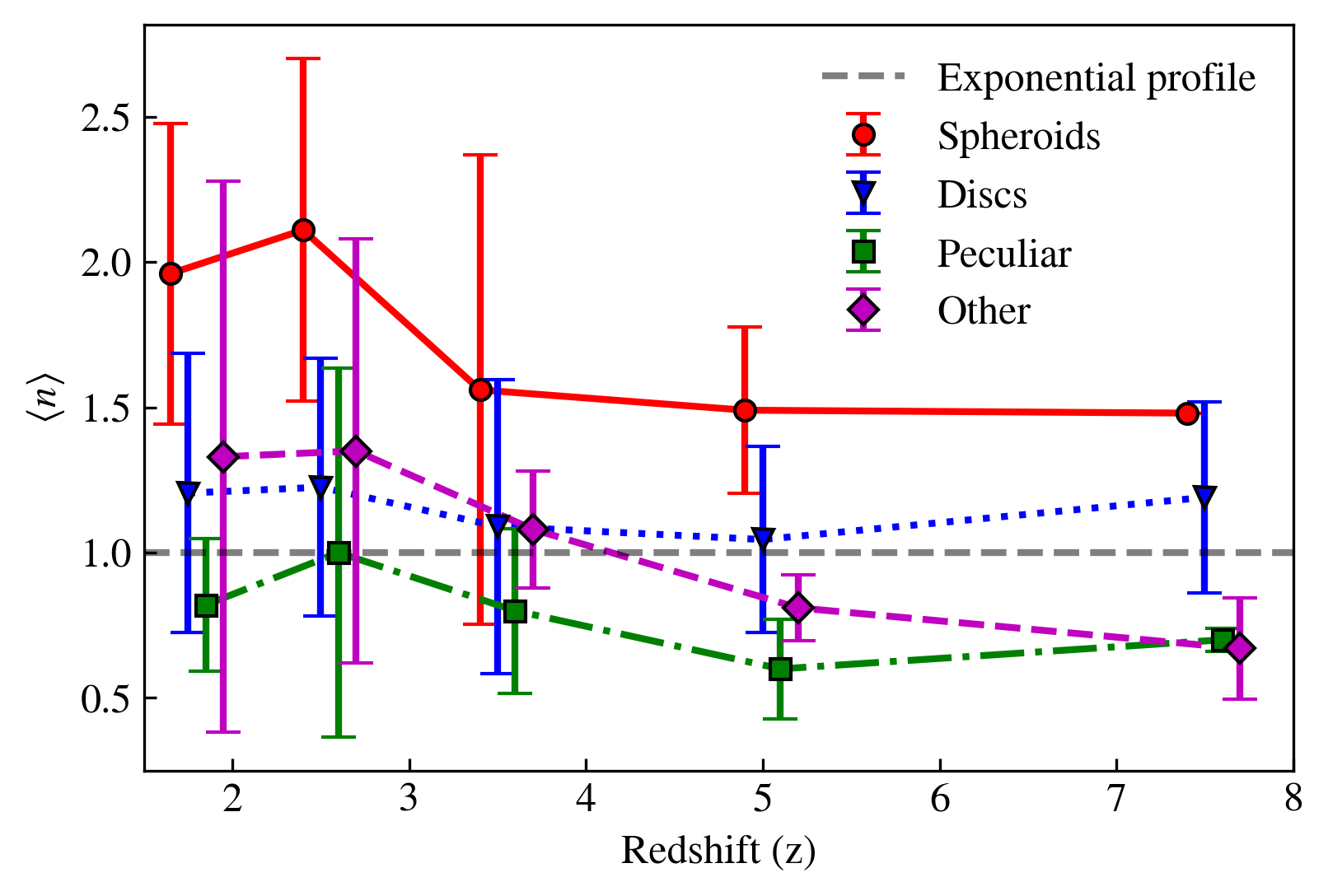}
        \caption{Sérsic index (n). The grey dashed line represents the $n = 1$ profile of an exponential disc.}
        \label{subfig:sersic_morph}
    \end{subfigure}
    
    \caption{Size and Sérsic index evolution as a function of visually determined morphology from \citet{Ferreira22b}. At all redshifts, we find that spheroid galaxies have the smallest radius, and the highest Sérsic index, displaying the compact, concentrated nature of these objects.}
    \label{fig: morph}
\end{figure}

For 470 galaxies within our sample, we use visual morphological classifications from \citet{Ferreira22b}, as defined in \autoref{subsec:visual} to determine how the properties and relations we have found are determined by overall morphology.  This is a small sample of galaxies, but represents one of the first times that we can examine these measured properties with visual morphologies.   Using these classifications, we present an analysis of size and Sérsic index as a function of morphology, as shown in \autoref{fig: morph}. 

\autoref{subfig:radius_morph} shows that for all galaxy types, radius decreases with increasing redshift, but at all redshifts, spheroid-type galaxies are the smallest. \autoref{subfig:sersic_morph} shows how Sérsic index varies with redshift for all galaxy types. Disc type galaxies have $n \sim 1$ as expected, with `peculiar' and `other' galaxies showing a slight decrease with redshift. Spheroid galaxies have the highest Sérsic index at all redshifts, significantly above other galaxy types, as found in \citet{Kartaltepe2022}. The small sizes of these galaxies combined with their high Sérsic index is a clear indicator of their compact, concentrated nature.

\subsection{Comparison with Image Simulations}

One major issue with a study such as this, which deals with imaging and the analysis of structure at galaxies at vastly different redshifts, is the fact that the surface brightness measurements of galaxies declines as $(1+z)^{4}$, and this can produce significant changes in the way that structure for distant galaxies would be imaged by a telescope. In fact, it is clear that galaxy structure can in principle change substantially and that many galaxies are potentially being missed at the highest redshifts \citep[][]{Conselice2003a,Whitney2020, Whitney2021}.  Therefore it is very important that we carry out simulations to determine if the trend we see in this paper, namely that galaxies get progressively smaller up to $z \sim 7$, is due to a real evolution or due to galaxies appearing smaller and fainter at higher redshifts.  Previous work using HST shows that whilst we are likely missing galaxies at the highest redshifts, we can still measure accurately their structural parameters \citep[][]{Whitney2020, Whitney2021}.

To understand this issue in depth for JWST data, we take a sample of 186 low-redshift galaxies at redshifts $0.5 < z < 1$, and create simulated images of these galaxies at higher redshifts, in intervals of $\Delta z = 0.5$, up to $ z = 7.5$, including all known cosmological effects. A representative sample of galaxies is used, with Sérsic indices ranging from $0.06 < n < 7.58$. This analysis is done in order to separate real evolution effects from redshift effects.
To do this simulation we use the redshifting code \textsc{AREIA}\footnote{\url{https://github.com/astroferreira/areia}}. We give here a brief overview of the steps taken are described below, for a more detailed discussion we refer the reader to \citet{tolhill21} and \citet{Whitney2021}. 

First, the source is extracted from the original stamp by measuring a segmentation map with \textsc{GalClean}\footnote{\url{https://github.com/astroferreira/galclean}}. Then, the image is geometrically re-binned from the source redshift to the target redshift based on the standard cosmology, preserving its flux. This is done to ensure that higher redshift sources have the appropriate geometric scaling due to the adopted cosmology. The resulting image from the rebinning has its flux scaled due to the cosmological dimming effect. Furthermore, shot noise is sampled from the source new light distribution, which is then convolved with the target JWST PSF of the rest-frame filter in the target redshift, to mimic an observation in the rest-frame optical. Then, the final redshifted source is placed on a random real CEERS background to mimic a real observation. This results in a sample of 2418 simulated images. 

Although \textsc{AREIA} allows the user to include size corrections and brightness corrections to mimic redshift evolution, we keep all intrinsic properties of the galaxies constant at each redshift, simulating only observational effects. 

Following the same method described in \autoref{subsec:pipeline}, we measure the sizes and Sérsic indices of our new simulated galaxies. We follow the same selection method, described in \autoref{sec:sample_selection} to ensure robust results. For our Sérsic index analysis, we further select objects with a \galfit\ measured magnitude $<$ 30.

\begin{figure}
    \centering
    \begin{subfigure}{\columnwidth}
        \includegraphics[width=\linewidth]{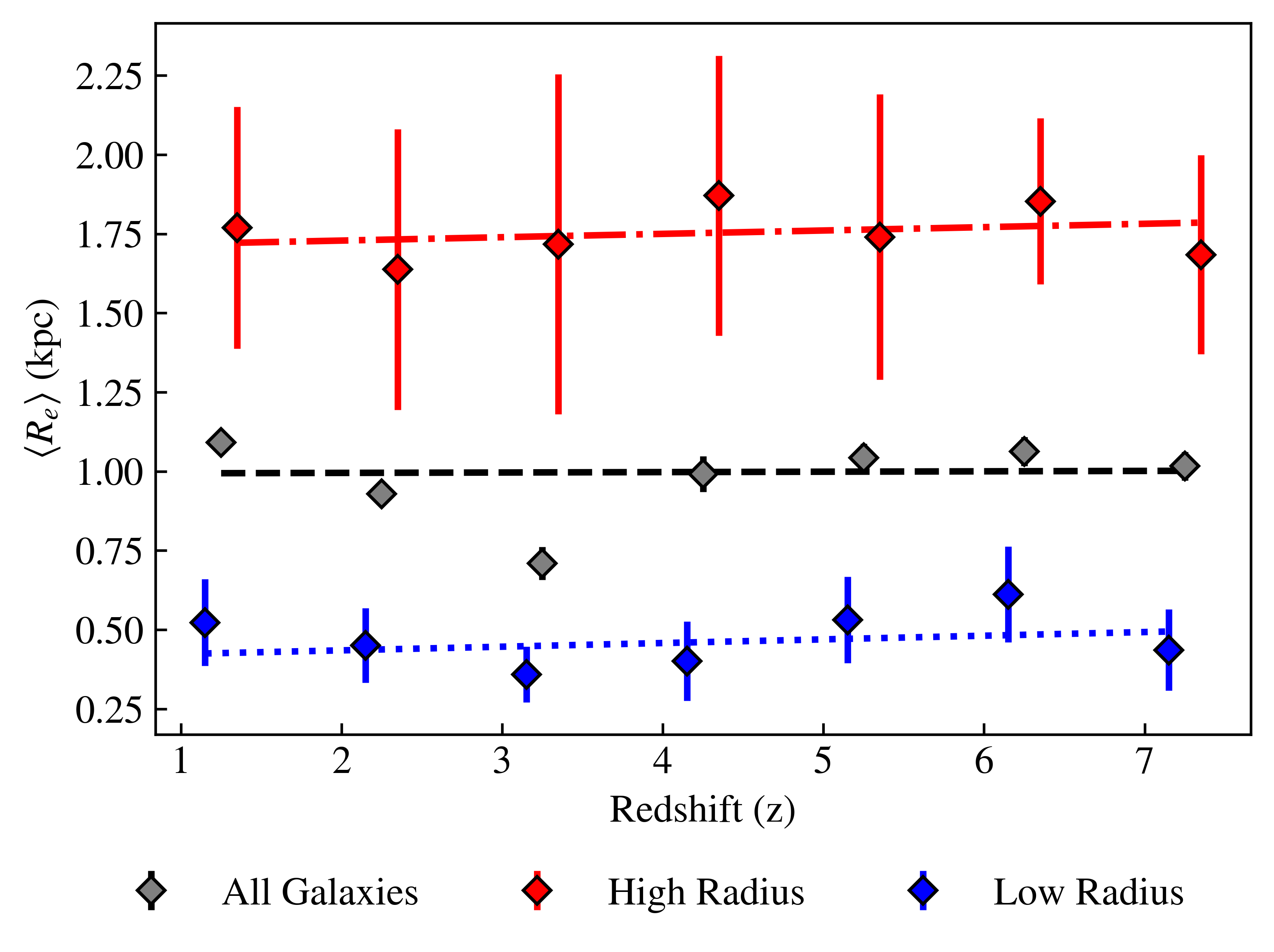}
        \caption{Median simulated galaxy sizes at each redshift. The gray points and line show this change for the simulated galaxies for the total simulated sample, whilst the red and blue are for those galaxies that are larger and smaller, respectively, than the median radius.}
        \label{fig:rad_sim}
    \end{subfigure}
    \begin{subfigure}{\columnwidth}
        \includegraphics[width=\linewidth]{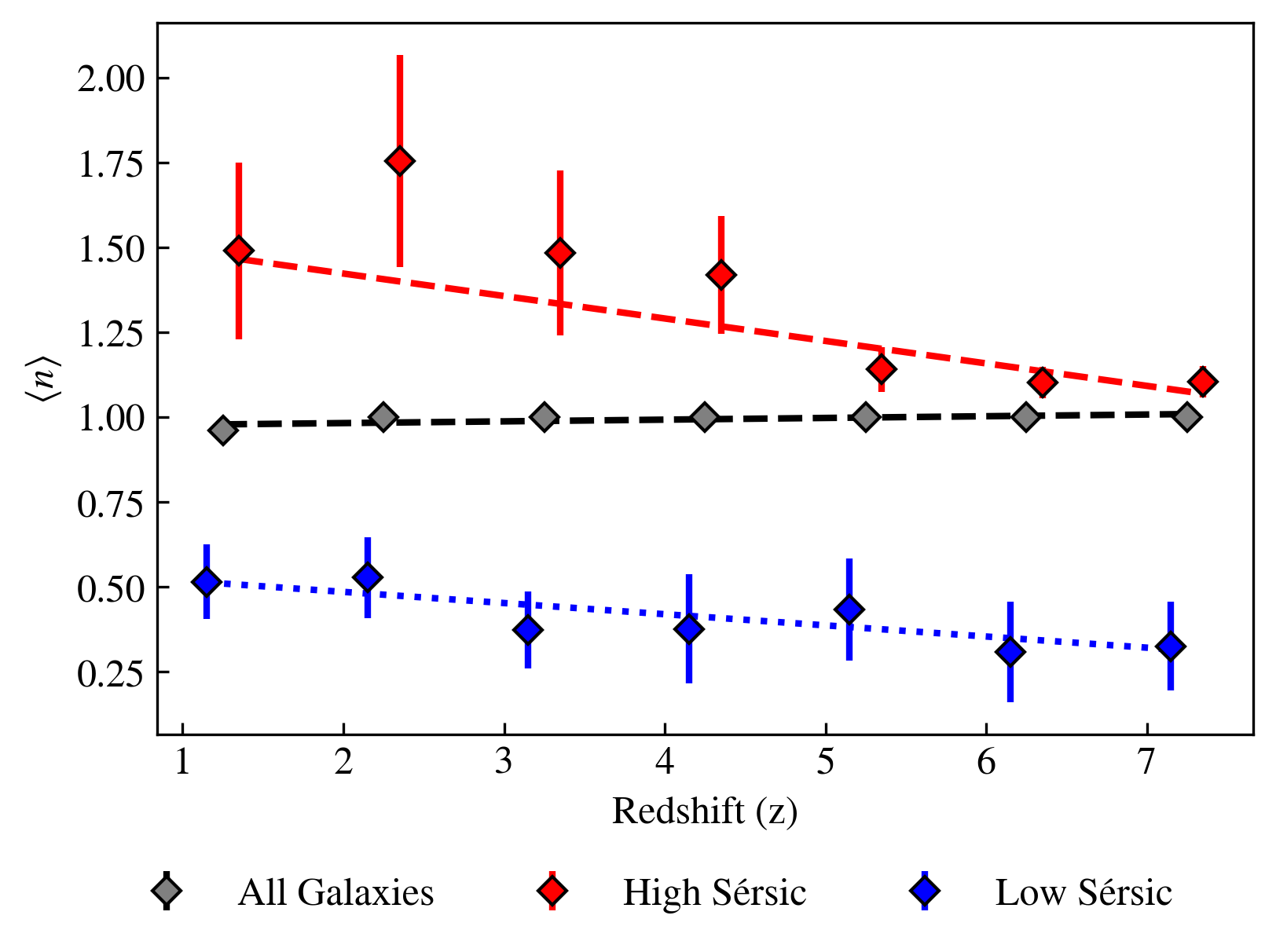}
        \caption{Median Sérsic index of simulated galaxies at each redshift. The lines are the same as explained in the plot of effective radius with redshift (\autoref{fig:rad_sim}).}
        \label{fig:sersic_sim}
    \end{subfigure}
    \caption{Sizes and Sérsic indices for our simulated galaxies in redshift bins. The error bars represent the standard error of the median for each bin. We divide these samples into different sub-types to determine how different selections would evolve differently within these simulations.}
    \label{fig:simulation}
\end{figure}
We also analyse the trends for objects with values above and below the overall median, indicated by the `high' and `low' radius and Sérsic groups.  

We use the \texttt{emcee} package to obtain the lines of best fit \citep{emcee}.
As shown in \autoref{fig:simulation}, we find that our sizes and Sérsic indices are best fit by linear fits, with gradients and errors stated in \autoref{tab:simulations}. We find our results are consistent with "flat slopes" - that is we find no change with redshift for the measured sizes and Sersic indices within these simulations.  This shows that the simulated galaxies continue to have very similar size measurements at different redshifts, meaning that in the absence of evolution we would expect the same galaxies to have the same effective radii and Sersic indices measured at all redshifts. This confirms that our method recovers the same result regardless of the redshift in which the galaxy is observed. 

This is vastly important for this work, and shows that the trends obtained in \autoref{sec:results}, are due to a change in galaxy properties and populations with increasing redshift, not because the galaxies appear smaller at higher redshifts due to cosmological or redshift effects.  

\begin{table}
    \centering
    \begin{subtable}{0.45\linewidth}
        \centering
        \begin{tabular}{c|c}
            \hline
            Radius Subset & Gradient \\
            \hline
            All Galaxies & $0.001\pm0.137$ \\
            High Radius & $0.009\pm0.159$ \\
            Low Radius & $0.010\pm0.138$ \\
            \hline
        \end{tabular}
        \caption{Radius (kpc)}
        \label{subtable1}
    \end{subtable}
    \hfill
    \begin{subtable}{0.45\linewidth}
        \centering
        \begin{tabular}{c|c}
            \hline
            Sérsic Subset & Gradient \\
            \hline
            All Galaxies & $0.005\pm0.139$ \\
            High Sérsic & $-0.067\pm0.148$ \\
            Low Sérsic & $-0.032\pm0.149$ \\
            \hline
        \end{tabular}
        \caption{Sérsic index (n)}
        \label{subtable2}
    \end{subtable}
    \caption{Best fit results for linear fits to all galaxies, and high and low subsets for both radius and Sérsic index. The slopes are consistent with being flat - with little change with redshift, showing that our main findings are due to evolutionary effects, not redshift effects.}
    \label{tab:simulations}
\end{table}

\section{Discussion}

The results in this paper reveal a myriad of new observational facts about the sizes and shape evolution of galaxies up to $z \sim 8$.   Our core result is that galaxies become progressively smaller at fixed stellar mass at higher redshifts. Whilst this was known for some time up to $z \sim 3$ \citep[][]{Trujillo2007, Buitrago2008, vanderwal2012, dokkum_2010}, \emph{JWST} is now allowing an analysis of this in the rest-frame optical light at higher redshifts than previously possible.  In fact, we find that from $z \sim 7$ to $z \sim 3$, galaxies with larger masses roughly double in size. This evolution is not as dramatic as is seen for the highest mass galaxies at $z < 3$ \citep[e.g.,][]{Buitrago2008} and we will require that more area be covered before we have statistics to probe these very high mass galaxies at such high redshift, as very few are imaged due to the limited numbers and sky coverage with existing and reliable JWST data.

Another major result is that we find very little variation in the size distribution and size evolution for our sample at $z > 3$ irrespective of how we divide the sample.  This is true for different Sérsic cuts, as well as for different cuts in the sSFR for these galaxies.   One of the main signatures of the formation of the Hubble sequence is a bifurcation in galaxies into morphologically distinct populations of star-forming and relatively passive systems (often simply divided into discs and ellipticals).  Whilst we are not seeing the entire formation of this Hubble sequence at $z \sim 3$, it is clear that the major bifurcation begins at this epoch and increasingly differentiates itself. This is likely due to different formation mechanisms coming into play at $z < 3$ that were not present at the higher redshifts.  This is likely something to do with mergers and feedback from either AGN or star formation \citep[e.g.,][]{Bluck2012}, particularly inside out star formation, where star forming galaxies begin to grow more rapidly than their passive counterparts. 

While we do not see much difference between galaxies at $z > 3$, we do find that galaxies have a well established difference in size and Sérsic index as a function of the wavelength of observation.   This is such that galaxies are more compact in redder wavelengths, showing that the outer parts are made up of more recent stars and star formation events.  This could be a sign that galaxies are forming inside-out and that we are witnessing the formation of bulges and the cores of giant galaxies at these early times that are growing outward from minor mergers and/or accreted gas in star formation \citep[e.g.,][]{tachella_15a, Tacchella_2015, Tacchella_2018, Nelson_2016, nelson_2019,Wilman_2020, Matharu_2022, roper_2023}.    This is consistent with the formation of bulges and disks occurring gradually at about $z \sim 3 $ and at lower redshifts \citep[e.g.,][]{margalef-bentabol2016}.

We also find that there is a correlation between stellar mass and size for galaxies -- for both galaxies that are relatively more star forming and those that are more passive -- up to $z \sim 3$.  Our JWST results allow us to accurately probe these mass ranges, even at the higher redshifts where HST has had a difficult time resolving these systems. These results provide another indication that something is regulating the sizes and masses of galaxies, and that this appears to be more present at $z < 3$ than at earlier times.  When we compare with simulations, such as the TNG50 simulation at $3 < z < 6$ (see \autoref{fig:rad_evolution}) we find that there is a good agreement. The causes of this change around $z \sim 3$, and the physics behind the mergers that produce these size increases within these simulations are still to be determined.

\subsection{Caveats and Future Work}
This study's limitations, which future research could address, include the relatively small number of galaxies in the higher redshift categories. This limitation stems from our reliance on catalogues developed from HST. While these catalogues offer accurate photometric redshifts for lower-z galaxies, they inevitably miss some fainter galaxies at higher redshifts. Future research that concentrates exclusively on high redshift galaxies, using JWST data, should not face this limitation. Additionally, as more deep, high-z observations are made, the growing number of known distant galaxies will further mitigate this issue.

We also note that our comparisons of the passive and star-forming populations in this work are relative to the average sSFR in each redshift bin. While this is clear throughout the analysis, we feel that this could be improved upon when greater populations of each are known, or with studies focusing entirely on quiescent or star forming galaxies, with stricter definitions of each. We also note that the average sSFR in each bin varies, but we chose this method as using a fixed value to define the populations in each redshift bin resulted in either the low-z or high-z bins being entirely one population, thus we chose to present a comparison of the more passive and more star forming population at each redshift. For example, using log$_{10}$(sSFR) $= - 9$ as a constant value resulted in the lowest redshift bin being made up of mostly passive systems in a way that did not allow for a comparison to be made.

\section{Summary and Conclusions}

We present an analysis of 1395 carefully selected massive galaxies with stellar masses\logm$>$9.5, within the CEERS and CANDELS fields between z = 0.5 and z = 8, using light profile fitting. Our galaxy sample is taken from the CANDELS field to enable the use of robust masses, redshifts, and star formation rates from optical to NIR data, which is necessary to obtain accurate measurements for galaxies over our large redshift range. In this paper we fit single Sérsic profiles to our galaxies and analyse their effective radius and Sérsic index to probe the evolution of these properties through cosmic time. We also probe the variation in size and shape as a dependence on stellar mass and wavelength. The trends we find are robust to redshift effects as we show through simulations of placing low redshift galaxies at high redshift that parameters would not change simply due to being at higher redshifts as imaged with JWST.

To carry out our analysis we use a custom built \galfit\ pipeline, fitting a single Sérsic fit, and fitting neighbouring galaxies where appropriate. We verify this method via a comparison with another galaxy profile fitting code, \imfit\, and find that our results are generally in good agreement between these two codes, and thus reliably measured. We then analyse the evolution of effective radius and Sérsic index, for our sample as a whole, and for quiescent/star-forming population, and high/low Sérsic index populations. We verify that our results are due to evolutionary effects rather than redshifting effects, by fitting the same galaxies redshifted to higher redshifts using the exact same process we apply on the real galaxies.  We find "flat" relations between measured parameters and redshift, thus confirming the robustness of our method and its ability to recover the correct measurements regardless of the distance to the objects. 

Our main findings are as follows:
\begin{itemize}
    \item Sérsic indices decrease on average with increasing redshift, suggesting a higher proportion of "disc-like" galaxies in the early universe. We find that passive and star-forming populations of galaxies show a different evolution of Sérsic index with redshift, with star-forming galaxies hovering around value of $n = 1$, suggesting that most star formation occurs within disc-like galaxies, at least in terms of structure. We cannot however rule out that some of these galaxies are involved in mergers.  In principle this confirms what has been found when classifying galaxies visually \citep[][]{Ferreira22b}.
    \item In general, more massive galaxies have a larger effective radius up to at least $z \sim 3$ compared to lower mass galaxies.  At redshifts higher than $z \sim 3$, at a fixed stellar mass, star-forming and quiescent galaxies have very similar sizes, suggesting that quiescent galaxies may not be smaller than star-forming galaxies at fixed stellar mass at high redshift, which has been a finding for almost 20 years at $z < 3$ \citep[e.g.,][]{Buitrago2008}.
    \item We find that galaxies appear more compact and smaller when observed in redder filters, and demonstrate the mass dependence of this effect, where more massive galaxies are more compact in redder filters than their lower mass counterparts. 
    \item We find that visually classified spheroid galaxies are smaller than other galaxy types at all redshifts, and have a higher Sérsic index at all redshifts than other galaxy types, showing their small, compact nature.
    \item We verify that our results are due to real evolutionary effects only, shown by fitting simulated high redshift galaxies with the intrinsic properties preserved, and we recover results that do not vary based on redshift effects.
\end{itemize}

Overall, this paper, we show that the evolution of galaxy size and structure continues to the highest redshifts, with disc-like galaxies forming most stars within the universe at all epochs. High redshift morphology studies are revealing a new picture of the structural evolution of galaxies, which will continue further with increasing numbers of high-redshift galaxies being discovered with JWST.

This study is just the start of this type of analysis.  The benefit of the CEERS fields is that we have very accurate photometric redshifts at lower redshifts, due to overlap with existing HST observations.  In the near future this will be available for many other and larger fields where more subtle changes in the size and structural features of galaxies will be studied, and this will lead to a more complete understanding of galaxy evolution over nearly the universe's entire history. Future studies with a larger sample of high redshift galaxies are needed to determine precisely when these aspects first formed, and future observations will provide these increased samples of high redshift galaxies, through upcoming wide and deep surveys. An alternative method of probing only the highest redshift galaxies could also be used, removing the reliance on HST detections.

\section*{Acknowledgements}
We thank the anonymous referee for their comments, which improved this paper.
QL, CC, JT, and NA acknowledge support from the ERC Advanced Investigator Grant EPOCHS (788113). DA and TH acknowledge support from STFC in the form of PhD studentships. LF acknowledges financial support from Coordenação de Aperfeiçoamento de Pessoal de Nível Superior - Brazil (CAPES) in the form of a PhD studentship.

This work is based on observations made with the NASA/ESA \textit{Hubble Space Telescope} (HST) and NASA/ESA/CSA \textit{James Webb Space Telescope} (JWST) obtained from the \texttt{Mikulski Archive for Space Telescopes} (\texttt{MAST}) at the \textit{Space Telescope Science Institute} (STScI), which is operated by the Association of Universities for Research in Astronomy, Inc., under NASA contract NAS 5-03127 for JWST, and NAS 5–26555 for HST. The authors thank all involved in the construction and operations of the telescope as well as those who designed and executed these observations.

The authors thank Anthony Holloway and Sotirios Sanidas for providing their expertise in high performance computing and other IT support throughout this work. This research made use of the following Python libraries: \textsc{Astropy} \citep{astropy2022}; \textsc{Morfometryka} \citep{ferrari2015}; \textsc{Pandas} \citep{pandas}; \textsc{Matplotlib} \citep{Hunter:2007}; \textsc{photutils} \citep{photutils}.

\section*{Data Availability}
The JWST/NIRCam images used in this work are available through the Mikulski Archive for Space  Telescopes (https://mast.stsci.edu/) under Proposal ID 1345. Additional data products will be made available upon reasonable request to the corresponding author.



\bibliographystyle{mnras}
\bibliography{paper} 

\begin{thebibliography}{}
\makeatletter
\relax
\def\mn@urlcharsother{\let\do\@makeother \do\$\do\&\do\#\do\^\do\_\do\%\do\~}
\def\mn@doi{\begingroup\mn@urlcharsother \@ifnextchar [ {\mn@doi@} {\mn@doi@[]}}
\def\mn@doi@[#1]#2{\def\@tempa{#1}\ifx\@tempa\@empty \href {http://dx.doi.org/#2} {doi:#2}\else \href {http://dx.doi.org/#2} {#1}\fi \endgroup}
\def\mn@eprint#1#2{\mn@eprint@#1:#2::\@nil}
\def\mn@eprint@arXiv#1{\href {http://arxiv.org/abs/#1} {{\tt arXiv:#1}}}
\def\mn@eprint@dblp#1{\href {http://dblp.uni-trier.de/rec/bibtex/#1.xml} {dblp:#1}}
\def\mn@eprint@#1:#2:#3:#4\@nil{\def\@tempa {#1}\def\@tempb {#2}\def\@tempc {#3}\ifx \@tempc \@empty \let \@tempc \@tempb \let \@tempb \@tempa \fi \ifx \@tempb \@empty \def\@tempb {arXiv}\fi \@ifundefined {mn@eprint@\@tempb}{\@tempb:\@tempc}{\expandafter \expandafter \csname mn@eprint@\@tempb\endcsname \expandafter{\@tempc}}}

\bibitem[\protect\citeauthoryear{Adams et~al.,}{Adams et~al.}{2023a}]{adams2023epochs}
Adams N.~J.,  et~al., 2023a, EPOCHS Paper II: The Ultraviolet Luminosity Function from $7.5<z<13.5$ using 110 square arcminutes of deep, blank-field data from the PEARLS Survey and Public Science Programmes (\mn@eprint {arXiv} {2304.13721})

\bibitem[\protect\citeauthoryear{{Adams} et~al.,}{{Adams} et~al.}{2023b}]{adams2023}
{Adams} N.~J.,  et~al., 2023b, \mn@doi [\mnras] {10.1093/mnras/stac3347}, \href {https://ui.adsabs.harvard.edu/abs/2023MNRAS.518.4755A} {518, 4755}

\bibitem[\protect\citeauthoryear{{Almosallam}, {Jarvis}  \& {Roberts}}{{Almosallam} et~al.}{2016}]{almosallam_2016}
{Almosallam} I.~A.,  {Jarvis} M.~J.,   {Roberts} S.~J.,  2016, \mn@doi [\mnras] {10.1093/mnras/stw1618}, \href {https://ui.adsabs.harvard.edu/abs/2016MNRAS.462..726A} {462, 726}

\bibitem[\protect\citeauthoryear{{Ashby} et~al.,}{{Ashby} et~al.}{2015}]{ashby2015}
{Ashby} M.~L.~N.,  et~al., 2015, \mn@doi [\apjs] {10.1088/0067-0049/218/2/33}, \href {https://ui.adsabs.harvard.edu/abs/2015ApJS..218...33A} {218, 33}

\bibitem[\protect\citeauthoryear{{Astropy Collaboration} et~al.,}{{Astropy Collaboration} et~al.}{2022}]{astropy2022}
{Astropy Collaboration} et~al., 2022, \mn@doi [\apj] {10.3847/1538-4357/ac7c74}, \href {https://ui.adsabs.harvard.edu/abs/2022ApJ...935..167A} {935, 167}

\bibitem[\protect\citeauthoryear{Atek et~al.,}{Atek et~al.}{2022}]{Atek_2022}
Atek H.,  et~al., 2022, \mn@doi [Monthly Notices of the Royal Astronomical Society] {10.1093/mnras/stac3144}, 519, 1201

\bibitem[\protect\citeauthoryear{{Austin} et~al.,}{{Austin} et~al.}{2023}]{austin_2023}
{Austin} D.,  et~al., 2023, \mn@doi [\apjl] {10.3847/2041-8213/ace18d}, \href {https://ui.adsabs.harvard.edu/abs/2023ApJ...952L...7A} {952, L7}

\bibitem[\protect\citeauthoryear{Bagley et~al.,}{Bagley et~al.}{2023}]{Bagley_2023}
Bagley M.~B.,  et~al., 2023, \mn@doi [The Astrophysical Journal Letters] {10.3847/2041-8213/acbb08}, 946, L12

\bibitem[\protect\citeauthoryear{{Bertin} \& {Arnouts}}{{Bertin} \& {Arnouts}}{1996}]{bertin1996}
{Bertin} E.,  {Arnouts} S.,  1996, \mn@doi [\aaps] {10.1051/aas:1996164}, \href {https://ui.adsabs.harvard.edu/abs/1996A&AS..117..393B} {117, 393}

\bibitem[\protect\citeauthoryear{{Bluck}, {Conselice}, {Buitrago}, {Gr{\"u}tzbauch}, {Hoyos}, {Mortlock}  \& {Bauer}}{{Bluck} et~al.}{2012}]{Bluck2012}
{Bluck} A. F.~L.,  {Conselice} C.~J.,  {Buitrago} F.,  {Gr{\"u}tzbauch} R.,  {Hoyos} C.,  {Mortlock} A.,   {Bauer} A.~E.,  2012, \mn@doi [\apj] {10.1088/0004-637X/747/1/34}, \href {https://ui.adsabs.harvard.edu/abs/2012ApJ...747...34B} {747, 34}

\bibitem[\protect\citeauthoryear{{Bradley} et~al.,}{{Bradley} et~al.}{2020}]{photutils}
{Bradley} L.,  et~al., 2020, {astropy/photutils: 1.0.0}, Zenodo, \mn@doi{10.5281/zenodo.4044744}

\bibitem[\protect\citeauthoryear{{Brammer}, {van Dokkum}  \& {Coppi}}{{Brammer} et~al.}{2008}]{brammer_2008}
{Brammer} G.~B.,  {van Dokkum} P.~G.,   {Coppi} P.,  2008, \mn@doi [\apj] {10.1086/591786}, \href {https://ui.adsabs.harvard.edu/abs/2008ApJ...686.1503B} {686, 1503}

\bibitem[\protect\citeauthoryear{{Bridge} et~al.,}{{Bridge} et~al.}{2019}]{bridge_2019}
{Bridge} J.~S.,  et~al., 2019, \mn@doi [\apj] {10.3847/1538-4357/ab3213}, \href {https://ui.adsabs.harvard.edu/abs/2019ApJ...882...42B} {882, 42}

\bibitem[\protect\citeauthoryear{{Bruzual} \& {Charlot}}{{Bruzual} \& {Charlot}}{2003}]{bruzual_2003}
{Bruzual} G.,  {Charlot} S.,  2003, \mn@doi [\mnras] {10.1046/j.1365-8711.2003.06897.x}, \href {https://ui.adsabs.harvard.edu/abs/2003MNRAS.344.1000B} {344, 1000}

\bibitem[\protect\citeauthoryear{Buitrago, Trujillo, Conselice, Bouwens, Dickinson  \& Yan}{Buitrago et~al.}{2008}]{Buitrago2008}
Buitrago F.,  Trujillo I.,  Conselice C.~J.,  Bouwens R.~J.,  Dickinson M.,   Yan H.,  2008, \mn@doi [The Astrophysical Journal] {10.1086/592836}, 687, L61

\bibitem[\protect\citeauthoryear{{Buitrago}, {Trujillo}, {Conselice}  \& {H{\"a}u{\ss}ler}}{{Buitrago} et~al.}{2013}]{Buitrago2013}
{Buitrago} F.,  {Trujillo} I.,  {Conselice} C.~J.,   {H{\"a}u{\ss}ler} B.,  2013, \mn@doi [\mnras] {10.1093/mnras/sts124}, \href {https://ui.adsabs.harvard.edu/abs/2013MNRAS.428.1460B} {428, 1460}

\bibitem[\protect\citeauthoryear{{Buitrago}, {Conselice}, {Epinat}, {Bedregal}, {Gr{\"u}tzbauch}  \& {Weiner}}{{Buitrago} et~al.}{2014}]{buitrago2014}
{Buitrago} F.,  {Conselice} C.~J.,  {Epinat} B.,  {Bedregal} A.~G.,  {Gr{\"u}tzbauch} R.,   {Weiner} B.~J.,  2014, \mn@doi [\mnras] {10.1093/mnras/stu034}, \href {https://ui.adsabs.harvard.edu/abs/2014MNRAS.439.1494B} {439, 1494}

\bibitem[\protect\citeauthoryear{{Caon}, {Capaccioli}  \& {D'Onofrio}}{{Caon} et~al.}{1993}]{caon_1993}
{Caon} N.,  {Capaccioli} M.,   {D'Onofrio} M.,  1993, \mn@doi [\mnras] {10.1093/mnras/265.4.1013}, \href {https://ui.adsabs.harvard.edu/abs/1993MNRAS.265.1013C} {265, 1013}

\bibitem[\protect\citeauthoryear{Castellano et~al.,}{Castellano et~al.}{2022}]{Castellano_2022}
Castellano M.,  et~al., 2022, \mn@doi [The Astrophysical Journal Letters] {10.3847/2041-8213/ac94d0}, 938, L15

\bibitem[\protect\citeauthoryear{{Chabrier}}{{Chabrier}}{2003}]{chabrier_2003}
{Chabrier} G.,  2003, \mn@doi [\pasp] {10.1086/376392}, \href {https://ui.adsabs.harvard.edu/abs/2003PASP..115..763C} {115, 763}

\bibitem[\protect\citeauthoryear{{Ciotti}}{{Ciotti}}{1991}]{ciotti_1991}
{Ciotti} L.,  1991, \aap, \href {https://ui.adsabs.harvard.edu/abs/1991A&A...249...99C} {249, 99}

\bibitem[\protect\citeauthoryear{{Ciotti} \& {Bertin}}{{Ciotti} \& {Bertin}}{1999}]{ciotti1999}
{Ciotti} L.,  {Bertin} G.,  1999, \mn@doi [\aap] {10.48550/arXiv.astro-ph/9911078}, \href {https://ui.adsabs.harvard.edu/abs/1999A&A...352..447C} {352, 447}

\bibitem[\protect\citeauthoryear{{Conselice}}{{Conselice}}{2003}]{Conselice2003a}
{Conselice} C.~J.,  2003, \mn@doi [\apjs] {10.1086/375001}, \href {https://ui.adsabs.harvard.edu/abs/2003ApJS..147....1C} {147, 1}

\bibitem[\protect\citeauthoryear{Conselice}{Conselice}{2014}]{Conselice2014a}
Conselice C.~J.,  2014, \mn@doi [Annual Review of Astronomy and Astrophysics] {10.1146/annurev-astro-081913-040037}, 52, 291

\bibitem[\protect\citeauthoryear{{Costantin} et~al.,}{{Costantin} et~al.}{2023}]{costantin2023}
{Costantin} L.,  et~al., 2023, \mn@doi [\apj] {10.3847/1538-4357/acb926}, \href {https://ui.adsabs.harvard.edu/abs/2023ApJ...946...71C} {946, 71}

\bibitem[\protect\citeauthoryear{{Curtis-Lake} et~al.,}{{Curtis-Lake} et~al.}{2023}]{curtis-lake_2023}
{Curtis-Lake} E.,  et~al., 2023, \mn@doi [Nature Astronomy] {10.1038/s41550-023-01918-w}, \href {https://ui.adsabs.harvard.edu/abs/2023NatAs...7..622C} {7, 622}

\bibitem[\protect\citeauthoryear{{Delgado-Serrano}, {Hammer}, {Yang}, {Puech}, {Flores}  \& {Rodrigues}}{{Delgado-Serrano} et~al.}{2010}]{Delgado-Serrano2010}
{Delgado-Serrano} R.,  {Hammer} F.,  {Yang} Y.~B.,  {Puech} M.,  {Flores} H.,   {Rodrigues} M.,  2010, \mn@doi [\aap] {10.1051/0004-6361/200912704}, \href {https://ui.adsabs.harvard.edu/abs/2010A&A...509A..78D} {509, A78}

\bibitem[\protect\citeauthoryear{Donnan et~al.,}{Donnan et~al.}{2022}]{Donnan_2022}
Donnan C.~T.,  et~al., 2022, \mn@doi [Monthly Notices of the Royal Astronomical Society] {10.1093/mnras/stac3472}, 518, 6011

\bibitem[\protect\citeauthoryear{Duncan et~al.,}{Duncan et~al.}{2014}]{duncan2014}
Duncan K.,  et~al., 2014, \mn@doi [Monthly Notices of the Royal Astronomical Society] {https://doi.org/10.1093/mnras/stu1622}, 444, 2960–2984

\bibitem[\protect\citeauthoryear{{Duncan} et~al.,}{{Duncan} et~al.}{2018a}]{duncan_2018a}
{Duncan} K.~J.,  et~al., 2018a, \mn@doi [\mnras] {10.1093/mnras/stx2536}, \href {https://ui.adsabs.harvard.edu/abs/2018MNRAS.473.2655D} {473, 2655}

\bibitem[\protect\citeauthoryear{{Duncan}, {Jarvis}, {Brown}  \& {R{\"o}ttgering}}{{Duncan} et~al.}{2018b}]{duncan_2018b}
{Duncan} K.~J.,  {Jarvis} M.~J.,  {Brown} M. J.~I.,   {R{\"o}ttgering} H. J.~A.,  2018b, \mn@doi [\mnras] {10.1093/mnras/sty940}, \href {https://ui.adsabs.harvard.edu/abs/2018MNRAS.477.5177D} {477, 5177}

\bibitem[\protect\citeauthoryear{{Duncan} et~al.,}{{Duncan} et~al.}{2019}]{duncan2019}
{Duncan} K.,  et~al., 2019, \mn@doi [\apj] {10.3847/1538-4357/ab148a}, \href {https://ui.adsabs.harvard.edu/abs/2019ApJ...876..110D} {876, 110}

\bibitem[\protect\citeauthoryear{{Erwin}}{{Erwin}}{2015}]{imfit}
{Erwin} P.,  2015, \mn@doi [\apj] {10.1088/0004-637X/799/2/226}, \href {https://ui.adsabs.harvard.edu/abs/2015ApJ...799..226E} {799, 226}

\bibitem[\protect\citeauthoryear{Ferrari, Carvalho  \& Trevisan}{Ferrari et~al.}{2015}]{ferrari2015}
Ferrari F.,  Carvalho R. R.~d.,   Trevisan M.,  2015, \mn@doi [The Astrophysical Journal] {https://doi.org/10.1088/0004-637x/814/1/55}, 814, 55

\bibitem[\protect\citeauthoryear{{Ferreira}, {Conselice}, {Duncan}, {Cheng}, {Griffiths}  \& {Whitney}}{{Ferreira} et~al.}{2020}]{Ferreira2020}
{Ferreira} L.,  {Conselice} C.~J.,  {Duncan} K.,  {Cheng} T.-Y.,  {Griffiths} A.,   {Whitney} A.,  2020, \mn@doi [\apj] {10.3847/1538-4357/ab8f9b}, \href {https://ui.adsabs.harvard.edu/abs/2020ApJ...895..115F} {895, 115}

\bibitem[\protect\citeauthoryear{{Ferreira} et~al.,}{{Ferreira} et~al.}{2022a}]{ferreira22a}
{Ferreira} L.,  et~al., 2022a, \mn@doi [arXiv e-prints] {10.48550/arXiv.2210.01110}, \href {https://ui.adsabs.harvard.edu/abs/2022arXiv221001110F} {p. arXiv:2210.01110}

\bibitem[\protect\citeauthoryear{{Ferreira} et~al.,}{{Ferreira} et~al.}{2022b}]{Ferreira22b}
{Ferreira} L.,  et~al., 2022b, \mn@doi [\apjl] {10.3847/2041-8213/ac947c}, \href {https://ui.adsabs.harvard.edu/abs/2022ApJ...938L...2F} {938, L2}

\bibitem[\protect\citeauthoryear{{Finkelstein} et~al.,}{{Finkelstein} et~al.}{2017}]{ceersprop}
{Finkelstein} S.~L.,  et~al., 2017, {The Cosmic Evolution Early Release Science (CEERS) Survey}, JWST Proposal ID 1345. Cycle 0 Early Release Science

\bibitem[\protect\citeauthoryear{{Finkelstein} et~al.,}{{Finkelstein} et~al.}{2022}]{ceers1}
{Finkelstein} S.~L.,  et~al., 2022, \mn@doi [arXiv e-prints] {10.48550/arXiv.2211.05792}, \href {https://ui.adsabs.harvard.edu/abs/2022arXiv221105792F} {p. arXiv:2211.05792}

\bibitem[\protect\citeauthoryear{Finkelstein et~al.,}{Finkelstein et~al.}{2023}]{Finkelstein_2023}
Finkelstein S.~L.,  et~al., 2023, \mn@doi [The Astrophysical Journal Letters] {10.3847/2041-8213/acade4}, 946, L13

\bibitem[\protect\citeauthoryear{Foreman-Mackey, Hogg, Lang  \& Goodman}{Foreman-Mackey et~al.}{2013}]{emcee}
Foreman-Mackey D.,  Hogg D.~W.,  Lang D.,   Goodman J.,  2013, \mn@doi [Publications of the Astronomical Society of the Pacific] {10.1086/670067}, 125, 306

\bibitem[\protect\citeauthoryear{{Furlong} et~al.,}{{Furlong} et~al.}{2017}]{furlong_2017}
{Furlong} M.,  et~al., 2017, \mn@doi [\mnras] {10.1093/mnras/stw2740}, \href {https://ui.adsabs.harvard.edu/abs/2017MNRAS.465..722F} {465, 722}

\bibitem[\protect\citeauthoryear{{Gaia Collaboration} et~al.,}{{Gaia Collaboration} et~al.}{2021}]{Gaia}
{Gaia Collaboration} et~al., 2021, \mn@doi [\aap] {10.1051/0004-6361/202039657}, \href {https://ui.adsabs.harvard.edu/abs/2021A&A...649A...1G} {649, A1}

\bibitem[\protect\citeauthoryear{{Grogin} et~al.,}{{Grogin} et~al.}{2011}]{CANDELS}
{Grogin} N.~A.,  et~al., 2011, \mn@doi [\apjs] {10.1088/0067-0049/197/2/35}, \href {https://ui.adsabs.harvard.edu/abs/2011ApJS..197...35G} {197, 35}

\bibitem[\protect\citeauthoryear{Harikane et~al.,}{Harikane et~al.}{2023}]{Harikane_2023}
Harikane Y.,  et~al., 2023, \mn@doi [The Astrophysical Journal Supplement Series] {10.3847/1538-4365/acaaa9}, 265, 5

\bibitem[\protect\citeauthoryear{Herschel}{Herschel}{1786}]{herschel_1786}
Herschel W.,  1786, Philosophical Transactions of the Royal Society of London, 76, 457–499

\bibitem[\protect\citeauthoryear{{Hoaglin}, {Mosteller}  \& {Tukey}}{{Hoaglin} et~al.}{1983}]{nmad}
{Hoaglin} D.~C.,  {Mosteller} F.,   {Tukey} J.~W.,  1983, {Understanding robust and exploratory data analysis}

\bibitem[\protect\citeauthoryear{{Hoyos} et~al.,}{{Hoyos} et~al.}{2011}]{hoyos2011}
{Hoyos} C.,  et~al., 2011, \mn@doi [\mnras] {10.1111/j.1365-2966.2010.17855.x}, \href {https://ui.adsabs.harvard.edu/abs/2011MNRAS.411.2439H} {411, 2439}

\bibitem[\protect\citeauthoryear{{Hoyos} et~al.,}{{Hoyos} et~al.}{2012}]{hoyos2012}
{Hoyos} C.,  et~al., 2012, \mn@doi [\mnras] {10.1111/j.1365-2966.2011.19918.x}, \href {https://ui.adsabs.harvard.edu/abs/2012MNRAS.419.2703H} {419, 2703}

\bibitem[\protect\citeauthoryear{{Hubble}}{{Hubble}}{1926}]{hubble1926}
{Hubble} E.~P.,  1926, \mn@doi [\apj] {10.1086/143018}, \href {https://ui.adsabs.harvard.edu/abs/1926ApJ....64..321H} {64, 321}

\bibitem[\protect\citeauthoryear{{Huertas-Company} et~al.,}{{Huertas-Company} et~al.}{2023}]{huertas_company_2023}
{Huertas-Company} M.,  et~al., 2023, \mn@doi [arXiv e-prints] {10.48550/arXiv.2305.02478}, \href {https://ui.adsabs.harvard.edu/abs/2023arXiv230502478H} {p. arXiv:2305.02478}

\bibitem[\protect\citeauthoryear{Hunter}{Hunter}{2007}]{Hunter:2007}
Hunter J.~D.,  2007, \mn@doi [Computing in Science & Engineering] {10.1109/MCSE.2007.55}, 9, 90

\bibitem[\protect\citeauthoryear{{Ito} et~al.,}{{Ito} et~al.}{2023}]{ito_2023}
{Ito} K.,  et~al., 2023, \mn@doi [arXiv e-prints] {10.48550/arXiv.2307.06994}, \href {https://ui.adsabs.harvard.edu/abs/2023arXiv230706994I} {p. arXiv:2307.06994}

\bibitem[\protect\citeauthoryear{{Jacobs} et~al.,}{{Jacobs} et~al.}{2023}]{jacobs2023}
{Jacobs} C.,  et~al., 2023, \mn@doi [\apjl] {10.3847/2041-8213/accd6d}, \href {https://ui.adsabs.harvard.edu/abs/2023ApJ...948L..13J} {948, L13}

\bibitem[\protect\citeauthoryear{{Kartaltepe} et~al.,}{{Kartaltepe} et~al.}{2022}]{Kartaltepe2022}
{Kartaltepe} J.~S.,  et~al., 2022, \mn@doi [arXiv e-prints] {10.48550/arXiv.2210.14713}, \href {https://ui.adsabs.harvard.edu/abs/2022arXiv221014713K} {p. arXiv:2210.14713}

\bibitem[\protect\citeauthoryear{Koekemoer et~al.,}{Koekemoer et~al.}{2011}]{koekemoer2011}
Koekemoer A.~M.,  et~al., 2011, \mn@doi [The Astrophysical Journal Supplement Series] {https://doi.org/10.1088/0067-0049/197/2/36}, 197, 36

\bibitem[\protect\citeauthoryear{{Kron}}{{Kron}}{1980}]{kron1980}
{Kron} R.~G.,  1980, \mn@doi [\apjs] {10.1086/190669}, \href {https://ui.adsabs.harvard.edu/abs/1980ApJS...43..305K} {43, 305}

\bibitem[\protect\citeauthoryear{{Kubo}, {Yamada}, {Ichikawa}, {Kajisawa}, {Matsuda}, {Tanaka}  \& {Umehata}}{{Kubo} et~al.}{2017}]{kubo_2017}
{Kubo} M.,  {Yamada} T.,  {Ichikawa} T.,  {Kajisawa} M.,  {Matsuda} Y.,  {Tanaka} I.,   {Umehata} H.,  2017, \mn@doi [\mnras] {10.1093/mnras/stx920}, \href {https://ui.adsabs.harvard.edu/abs/2017MNRAS.469.2235K} {469, 2235}

\bibitem[\protect\citeauthoryear{{Lotz}, {Primack}  \& {Madau}}{{Lotz} et~al.}{2004}]{Lotz2004}
{Lotz} J.~M.,  {Primack} J.,   {Madau} P.,  2004, \mn@doi [\aj] {10.1086/421849}, \href {https://ui.adsabs.harvard.edu/abs/2004AJ....128..163L} {128, 163}

\bibitem[\protect\citeauthoryear{{Lovell} et~al.,}{{Lovell} et~al.}{2023}]{lovell_2023}
{Lovell} C.~C.,  et~al., 2023, \mn@doi [\mnras] {10.1093/mnras/stad2550}, \href {https://ui.adsabs.harvard.edu/abs/2023MNRAS.tmp.2454L} {}

\bibitem[\protect\citeauthoryear{{Ma} et~al.,}{{Ma} et~al.}{2018}]{ma_2018}
{Ma} X.,  et~al., 2018, \mn@doi [\mnras] {10.1093/mnras/sty684}, \href {https://ui.adsabs.harvard.edu/abs/2018MNRAS.477..219M} {477, 219}

\bibitem[\protect\citeauthoryear{{Mager}, {Conselice}, {Seibert}, {Gusbar}, {Katona}, {Villari}, {Madore}  \& {Windhorst}}{{Mager} et~al.}{2018}]{Mager2018}
{Mager} V.~A.,  {Conselice} C.~J.,  {Seibert} M.,  {Gusbar} C.,  {Katona} A.~P.,  {Villari} J.~M.,  {Madore} B.~F.,   {Windhorst} R.~A.,  2018, \mn@doi [\apj] {10.3847/1538-4357/aad59e}, \href {https://ui.adsabs.harvard.edu/abs/2018ApJ...864..123M} {864, 123}

\bibitem[\protect\citeauthoryear{{Margalef-Bentabol}, {Conselice}, {Mortlock}, {Hartley}, {Duncan}, {Ferguson}, {Dekel}  \& {Primack}}{{Margalef-Bentabol} et~al.}{2016}]{margalef-bentabol2016}
{Margalef-Bentabol} B.,  {Conselice} C.~J.,  {Mortlock} A.,  {Hartley} W.,  {Duncan} K.,  {Ferguson} H.~C.,  {Dekel} A.,   {Primack} J.~R.,  2016, \mn@doi [\mnras] {10.1093/mnras/stw1451}, \href {https://ui.adsabs.harvard.edu/abs/2016MNRAS.461.2728M} {461, 2728}

\bibitem[\protect\citeauthoryear{{Marshall}, {Wilkins}, {Di Matteo}, {Roper}, {Vijayan}, {Ni}, {Feng}  \& {Croft}}{{Marshall} et~al.}{2022}]{marshall_2022}
{Marshall} M.~A.,  {Wilkins} S.,  {Di Matteo} T.,  {Roper} W.~J.,  {Vijayan} A.~P.,  {Ni} Y.,  {Feng} Y.,   {Croft} R. A.~C.,  2022, \mn@doi [\mnras] {10.1093/mnras/stac380}, \href {https://ui.adsabs.harvard.edu/abs/2022MNRAS.511.5475M} {511, 5475}

\bibitem[\protect\citeauthoryear{Matharu et~al.,}{Matharu et~al.}{2022}]{Matharu_2022}
Matharu J.,  et~al., 2022, \mn@doi [The Astrophysical Journal] {10.3847/1538-4357/ac8471}, 937, 16

\bibitem[\protect\citeauthoryear{Morishita et~al.,}{Morishita et~al.}{2023}]{morishita2023}
Morishita T.,  et~al., 2023, Enhanced Sub-kpc Scale Star-formation: Results From A JWST Size Analysis of 339 Galaxies At 5<z<14 (\mn@eprint {arXiv} {2308.05018})

\bibitem[\protect\citeauthoryear{{Mortlock} et~al.,}{{Mortlock} et~al.}{2013}]{Mortlock2013}
{Mortlock} A.,  et~al., 2013, \mn@doi [\mnras] {10.1093/mnras/stt793}, \href {https://ui.adsabs.harvard.edu/abs/2013MNRAS.433.1185M} {433, 1185}

\bibitem[\protect\citeauthoryear{Naidu et~al.,}{Naidu et~al.}{2022}]{Naidu_2022}
Naidu R.~P.,  et~al., 2022, \mn@doi [The Astrophysical Journal Letters] {10.3847/2041-8213/ac9b22}, 940, L14

\bibitem[\protect\citeauthoryear{Nelson et~al.,}{Nelson et~al.}{2016}]{Nelson_2016}
Nelson E.~J.,  et~al., 2016, \mn@doi [The Astrophysical Journal] {10.3847/0004-637X/828/1/27}, 828, 27

\bibitem[\protect\citeauthoryear{{Nelson} et~al.,}{{Nelson} et~al.}{2019}]{nelson_2019}
{Nelson} E.~J.,  et~al., 2019, \mn@doi [\apj] {10.3847/1538-4357/aaf38a}, \href {https://ui.adsabs.harvard.edu/abs/2019ApJ...870..130N} {870, 130}

\bibitem[\protect\citeauthoryear{{Oke}}{{Oke}}{1974}]{oke}
{Oke} J.~B.,  1974, \mn@doi [\apjs] {10.1086/190287}, \href {https://ui.adsabs.harvard.edu/abs/1974ApJS...27...21O} {27, 21}

\bibitem[\protect\citeauthoryear{{Oke} \& {Gunn}}{{Oke} \& {Gunn}}{1983}]{oke_gunn}
{Oke} J.~B.,  {Gunn} J.~E.,  1983, \mn@doi [\apj] {10.1086/160817}, \href {https://ui.adsabs.harvard.edu/abs/1983ApJ...266..713O} {266, 713}

\bibitem[\protect\citeauthoryear{{Ono} et~al.,}{{Ono} et~al.}{2023}]{ono_2023}
{Ono} Y.,  et~al., 2023, \mn@doi [\apj] {10.3847/1538-4357/acd44a}, \href {https://ui.adsabs.harvard.edu/abs/2023ApJ...951...72O} {951, 72}

\bibitem[\protect\citeauthoryear{{Papaderos}, {{\"O}stlin}  \& {Breda}}{{Papaderos} et~al.}{2023}]{papaderos2023}
{Papaderos} P.,  {{\"O}stlin} G.,   {Breda} I.,  2023, \mn@doi [\aap] {10.1051/0004-6361/202245769}, \href {https://ui.adsabs.harvard.edu/abs/2023A&A...673A..30P} {673, A30}

\bibitem[\protect\citeauthoryear{Paulino-Afonso, Sobral, Buitrago  \& Afonso}{Paulino-Afonso et~al.}{2016}]{paulino-afonso}
Paulino-Afonso A.,  Sobral D.,  Buitrago F.,   Afonso J.,  2016, \mn@doi [Monthly Notices of the Royal Astronomical Society] {10.1093/mnras/stw2933}, 465, 2717

\bibitem[\protect\citeauthoryear{{Peng}, {Ho}, {Impey}  \& {Rix}}{{Peng} et~al.}{2002}]{Galfit1}
{Peng} C.~Y.,  {Ho} L.~C.,  {Impey} C.~D.,   {Rix} H.-W.,  2002, \mn@doi [\aj] {10.1086/340952}, \href {https://ui.adsabs.harvard.edu/abs/2002AJ....124..266P} {124, 266}

\bibitem[\protect\citeauthoryear{{Peng}, {Ho}, {Impey}  \& {Rix}}{{Peng} et~al.}{2010}]{Galfit2}
{Peng} C.~Y.,  {Ho} L.~C.,  {Impey} C.~D.,   {Rix} H.-W.,  2010, \mn@doi [\aj] {10.1088/0004-6256/139/6/2097}, \href {https://ui.adsabs.harvard.edu/abs/2010AJ....139.2097P} {139, 2097}

\bibitem[\protect\citeauthoryear{{Perrin}, {Sivaramakrishnan}, {Lajoie}, {Elliott}, {Pueyo}, {Ravindranath}  \& {Albert}}{{Perrin} et~al.}{2014}]{perrin2014}
{Perrin} M.~D.,  {Sivaramakrishnan} A.,  {Lajoie} C.-P.,  {Elliott} E.,  {Pueyo} L.,  {Ravindranath} S.,   {Albert} L.,  2014, in {Oschmann} Jacobus~M. J.,  {Clampin} M.,  {Fazio} G.~G.,   {MacEwen} H.~A.,  eds,  Society of Photo-Optical Instrumentation Engineers (SPIE) Conference Series Vol. 9143, Space Telescopes and Instrumentation 2014: Optical, Infrared, and Millimeter Wave. p. 91433X, \mn@doi{10.1117/12.2056689}

\bibitem[\protect\citeauthoryear{{Popping} et~al.,}{{Popping} et~al.}{2022}]{popping_2022}
{Popping} G.,  et~al., 2022, \mn@doi [\mnras] {10.1093/mnras/stab3312}, \href {https://ui.adsabs.harvard.edu/abs/2022MNRAS.510.3321P} {510, 3321}

\bibitem[\protect\citeauthoryear{{Reback} et~al.,}{{Reback} et~al.}{2022}]{pandas}
{Reback} J.,  et~al., 2022, {pandas-dev/pandas: Pandas 1.4.2}, Zenodo, \mn@doi{10.5281/zenodo.3509134}

\bibitem[\protect\citeauthoryear{Rieke et~al.,}{Rieke et~al.}{2022}]{nircam}
Rieke M.~J.,  et~al., 2022, NIRCam Performance on JWST In Flight (\mn@eprint {arXiv} {2212.12069})

\bibitem[\protect\citeauthoryear{{Robertson} et~al.,}{{Robertson} et~al.}{2023}]{robertson_2023}
{Robertson} B.~E.,  et~al., 2023, \mn@doi [\apjl] {10.3847/2041-8213/aca086}, \href {https://ui.adsabs.harvard.edu/abs/2023ApJ...942L..42R} {942, L42}

\bibitem[\protect\citeauthoryear{Roper, Lovell, Vijayan, Marshall, Irodotou, Kuusisto, Thomas  \& Wilkins}{Roper et~al.}{2022}]{Roper_2022}
Roper W.~J.,  Lovell C.~C.,  Vijayan A.~P.,  Marshall M.~A.,  Irodotou D.,  Kuusisto J.~K.,  Thomas P.~A.,   Wilkins S.~M.,  2022, \mn@doi [Monthly Notices of the Royal Astronomical Society] {10.1093/mnras/stac1368}, 514, 1921

\bibitem[\protect\citeauthoryear{{Roper} et~al.,}{{Roper} et~al.}{2023}]{roper_2023}
{Roper} W.~J.,  et~al., 2023, \mn@doi [arXiv e-prints] {10.48550/arXiv.2301.05228}, \href {https://ui.adsabs.harvard.edu/abs/2023arXiv230105228R} {p. arXiv:2301.05228}

\bibitem[\protect\citeauthoryear{Rosse}{Rosse}{1850}]{rosse_1850}
Rosse T. E.~o.,  1850, Philosophical Transactions of the Royal Society of London, 140, 499–514

\bibitem[\protect\citeauthoryear{{Santini} et~al.,}{{Santini} et~al.}{2015}]{santini_2015}
{Santini} P.,  et~al., 2015, \mn@doi [\apj] {10.1088/0004-637X/801/2/97}, \href {https://ui.adsabs.harvard.edu/abs/2015ApJ...801...97S} {801, 97}

\bibitem[\protect\citeauthoryear{{Schawinski} et~al.,}{{Schawinski} et~al.}{2014}]{Schawinski2014}
{Schawinski} K.,  et~al., 2014, \mn@doi [\mnras] {10.1093/mnras/stu327}, \href {https://ui.adsabs.harvard.edu/abs/2014MNRAS.440..889S} {440, 889}

\bibitem[\protect\citeauthoryear{{S{\'e}rsic}}{{S{\'e}rsic}}{1963}]{Sersic}
{S{\'e}rsic} J.~L.,  1963, Boletin de la Asociacion Argentina de Astronomia La Plata Argentina, \href {https://ui.adsabs.harvard.edu/abs/1963BAAA....6...41S} {6, 41}

\bibitem[\protect\citeauthoryear{Stefanon et~al.,}{Stefanon et~al.}{2017}]{stefanon2017}
Stefanon M.,  et~al., 2017, \mn@doi [The Astrophysical Journal Supplement Series] {https://doi.org/10.3847/1538-4365/aa66cb}, 229, 32

\bibitem[\protect\citeauthoryear{{Suess} et~al.,}{{Suess} et~al.}{2022}]{Suess2022}
{Suess} K.~A.,  et~al., 2022, \mn@doi [\apjl] {10.3847/2041-8213/ac8e06}, \href {https://ui.adsabs.harvard.edu/abs/2022ApJ...937L..33S} {937, L33}

\bibitem[\protect\citeauthoryear{Tacchella et~al.,}{Tacchella et~al.}{2015a}]{tachella_15a}
Tacchella S.,  et~al., 2015a, \mn@doi [Science] {10.1126/science.1261094}, 348, 314

\bibitem[\protect\citeauthoryear{Tacchella et~al.,}{Tacchella et~al.}{2015b}]{Tacchella_2015}
Tacchella S.,  et~al., 2015b, \mn@doi [The Astrophysical Journal] {10.1088/0004-637X/802/2/101}, 802, 101

\bibitem[\protect\citeauthoryear{Tacchella et~al.,}{Tacchella et~al.}{2018}]{Tacchella_2018}
Tacchella S.,  et~al., 2018, \mn@doi [The Astrophysical Journal] {10.3847/1538-4357/aabf8b}, 859, 56

\bibitem[\protect\citeauthoryear{{Tacchella} et~al.,}{{Tacchella} et~al.}{2023}]{tacchella_2023}
{Tacchella} S.,  et~al., 2023, \mn@doi [\apj] {10.3847/1538-4357/acdbc6}, \href {https://ui.adsabs.harvard.edu/abs/2023ApJ...952...74T} {952, 74}

\bibitem[\protect\citeauthoryear{{Taylor-Mager}, {Conselice}, {Windhorst}  \& {Jansen}}{{Taylor-Mager} et~al.}{2007}]{TaylorMager2007}
{Taylor-Mager} V.~A.,  {Conselice} C.~J.,  {Windhorst} R.~A.,   {Jansen} R.~A.,  2007, \mn@doi [\apj] {10.1086/511806}, \href {https://ui.adsabs.harvard.edu/abs/2007ApJ...659..162T} {659, 162}

\bibitem[\protect\citeauthoryear{{Tohill}, {Ferreira}, {Conselice}, {Bamford}  \& {Ferrari}}{{Tohill} et~al.}{2021}]{tolhill21}
{Tohill} C.,  {Ferreira} L.,  {Conselice} C.~J.,  {Bamford} S.~P.,   {Ferrari} F.,  2021, \mn@doi [\apj] {10.3847/1538-4357/ac033c}, \href {https://ui.adsabs.harvard.edu/abs/2021ApJ...916....4T} {916, 4}

\bibitem[\protect\citeauthoryear{Trujillo, Conselice, Bundy, Cooper, Eisenhardt  \& Ellis}{Trujillo et~al.}{2007}]{Trujillo2007}
Trujillo I.,  Conselice C.~J.,  Bundy K.,  Cooper M.~C.,  Eisenhardt P.,   Ellis R.~S.,  2007, \mn@doi [Monthly Notices of the Royal Astronomical Society] {10.1111/j.1365-2966.2007.12388.x}, 382, 109

\bibitem[\protect\citeauthoryear{{Whitney}, {Conselice}, {Duncan}  \& {Spitler}}{{Whitney} et~al.}{2020}]{Whitney2020}
{Whitney} A.,  {Conselice} C.~J.,  {Duncan} K.,   {Spitler} L.~R.,  2020, \mn@doi [\apj] {10.3847/1538-4357/abb824}, \href {https://ui.adsabs.harvard.edu/abs/2020ApJ...903...14W} {903, 14}

\bibitem[\protect\citeauthoryear{Whitney, Ferreira, Conselice  \& Duncan}{Whitney et~al.}{2021}]{Whitney2021}
Whitney A.,  Ferreira L.,  Conselice C.~J.,   Duncan K.,  2021, \mn@doi [The Astrophysical Journal] {10.3847/1538-4357/ac1422}, 919, 139

\bibitem[\protect\citeauthoryear{Wilman et~al.,}{Wilman et~al.}{2020}]{Wilman_2020}
Wilman D.~J.,  et~al., 2020, \mn@doi [The Astrophysical Journal] {10.3847/1538-4357/ab7914}, 892, 1

\bibitem[\protect\citeauthoryear{{Windhorst} et~al.,}{{Windhorst} et~al.}{2002}]{Windhorst2002}
{Windhorst} R.~A.,  et~al., 2002, \mn@doi [\apjs] {10.1086/341556}, \href {https://ui.adsabs.harvard.edu/abs/2002ApJS..143..113W} {143, 113}

\bibitem[\protect\citeauthoryear{Yan, Ma, Ling, Cheng  \& Huang}{Yan et~al.}{2022}]{Yan_2022}
Yan H.,  Ma Z.,  Ling C.,  Cheng C.,   Huang J.-S.,  2022, \mn@doi [The Astrophysical Journal Letters] {10.3847/2041-8213/aca80c}, 942, L9

\bibitem[\protect\citeauthoryear{{Yang} et~al.,}{{Yang} et~al.}{2022}]{Yang2022}
{Yang} L.,  et~al., 2022, \mn@doi [\apjl] {10.3847/2041-8213/ac8803}, \href {https://ui.adsabs.harvard.edu/abs/2022ApJ...938L..17Y} {938, L17}

\bibitem[\protect\citeauthoryear{{van Dokkum} et~al.,}{{van Dokkum} et~al.}{2010}]{dokkum_2010}
{van Dokkum} P.~G.,  et~al., 2010, \mn@doi [\apj] {10.1088/0004-637X/709/2/1018}, \href {https://ui.adsabs.harvard.edu/abs/2010ApJ...709.1018V} {709, 1018}

\bibitem[\protect\citeauthoryear{{van der Wel} et~al.,}{{van der Wel} et~al.}{2012}]{vanderwal2012}
{van der Wel} A.,  et~al., 2012, \mn@doi [\apjs] {10.1088/0067-0049/203/2/24}, \href {https://ui.adsabs.harvard.edu/abs/2012ApJS..203...24V} {203, 24}

\bibitem[\protect\citeauthoryear{{van der Wel} et~al.,}{{van der Wel} et~al.}{2014}]{vdw2014}
{van der Wel} A.,  et~al., 2014, \mn@doi [\apj] {10.1088/0004-637X/788/1/28}, \href {https://ui.adsabs.harvard.edu/abs/2014ApJ...788...28V} {788, 28}

\makeatother
\end{thebibliography}





\bsp	
\label{lastpage}
\end{document}